\documentclass{aa}
\usepackage{epsf}

\newcommand{\iraf}{{\sc iraf}}
\newcommand{\midas}{{\sc midas}}
\newcommand{\timmi}{{\sc timmi}}

\newcommand{\irspec}{{\sc irspec}}
\newcommand{\caspir}{{\sc caspir}}
\newcommand{\class}{{\sc class}}

\newcommand{\iras}{{\sl IRAS}}
\newcommand{\hst}{{\sl HST}}
\newcommand{\usno}{{\sl USNO}}
\newcommand{\gsc}{{\sl GSC}}
\newcommand{\denis}{{\sl DENIS}}
\newcommand{\eso}{{\sl ESO}}
\newcommand{\usnof}{{\sl USNO--A2.0}}
\newcommand{\gscf}{{\sl GSC\,1.1}}
\newcommand{\denisf}{{\sl DENIS}}
\newcommand{\usnos}{{\sl U}}
\newcommand{\gscs}{{\sl G}}
\newcommand{\deniss}{{\sl D}}

\newcommand{\kms}{km\,s$^{-1}$}

\newcommand{\brg}{Br$\gamma$}
\newcommand{\jv}{{\sl V}}
\newcommand{\jr}{{\sl R}}
\newcommand{\jj}{{\sl J}}
\newcommand{\jh}{{\sl H}}
\newcommand{\jk}{{\sl K}}
\newcommand{\jl}{{\sl L}}
\newcommand{\jn}{{\sl N}}
\newcommand{\jhk}{{\jj \jh \jk}}
\newcommand{\jhkl}{{\jj \jh \jk \jl}}

\newcommand{\h}{\rlap{:}}

\newcommand{\f}{\phantom{0}}

\newcommand{\ctr}[3]{\multicolumn{#1}{c}{#2\hspace*{#3mm}}}

\def\ntd#1{\vtop{\footnotesize\hsize=\textwidth\leavevmode#1\hspace*{\fill}}}
\begin{document}
\thesaurus{08(08.16.4; 08.03.4; 08.05.3; 09.04.3; 09.16.1; 13.09.6)}
\title{Near-infrared and Br$\gamma$ observations of post-AGB stars\thanks{%
Based on observations made at the European Southern Observatory, La
Silla, Chile; the Australia Telescope Compact Array, which is 
funded by the Commonwealth of Australia for
operations as a National Facility managed by CSIRO;
Siding Spring Observatory.}}
\author{G.C. Van de Steene\inst{1,2}
\and 
P.A.M. van Hoof\inst{3,4,}\thanks{Currently staying at CITA, Toronto.}
\and
P. R. Wood\inst{1}
}
\offprints{G.C. Van de Steene, ROB, Ringlaan~3, B-1180 Brussels, Belgium}
\institute{Research School of Astronomy and Astrophysics,
Mount Stromlo Observatory, Private Bag, Weston Creek, ACT 2611, Australia
\and
European Southern Observatory, Casilla 19001, 19 Santiago, Chile
\and
Kapteyn Astronomical Institute, P.O. Box 800, 9700 AV Groningen,
The Netherlands
\and
University of Kentucky, Department of Physics and Astronomy, 177 CP Building, Lexington, KY 40506--0055, USA}
\date{Received date; accepted date}
\maketitle

\begin{abstract}

In this article we report further investigations of the \iras\ selected
sample of Planetary Nebula (PN) candidates that was presented in Van
de Steene \& Pottasch (\cite{VdSteene93}). About 20~\% of the
candidates in that sample have been detected in the radio and/or
H$\alpha$ and later confirmed as PNe. Here we investigate the infrared 
properties of the \iras\ sources not confirmed as PNe.

We observed 28 objects in the \jn-band of which 20 were detected and 5 were
resolved, despite adverse weather conditions. We obtained medium resolution
\brg\ spectra and we took high resolution \jhkl\ images of these 20 objects.
We critically assessed the identification of the \iras\ counterpart in the
images and compared our identification with others in the literature. 
High spatial resolution and a telescope with very accurate pointing are
crucial for correct identification of the \iras\ counterparts in these crowded
fields.

Of sixteen positively identified objects, seven show \brg\ in
absorption. The absorption lines are very narrow in six objects,
indicating a low surface gravity. Another six objects show \brg\ in
emission. Two of these also show photospheric absorption lines. All
emission line sources have a strong underlying continuum, unlike
normal PNe. In another three objects, no clear
\brg\ absorption or emission was visible. 
The fact that our objects were mostly selected from the region in
the \iras\ color-color diagram where typically PNe are found, may explain our
higher detection rate of emission line objects compared to previous studies,
which selected their candidates from a region between AGB and PNe. 

The objects showing \brg\ in emission were re-observed in the radio
continuum with the Australia Telescope Compact Array. None of them
were detected above a detection limit of 0.55~mJy/beam at 6~cm and
0.7~mJy/beam at 3~cm, while they should have been easily detected if
the radio flux was optically thin and Case~B recombination was
applicable. It is suggested that the \brg\ emission originates in the
post-Asymptotic Giant Branch (post-AGB) wind, and that the central
star is not yet hot enough to ionize the AGB shell.

We measured the \jhkl\ magnitudes of the objects and present their
infrared spectral energy distributions. They are typical for post-AGB
stars according to the scheme of van der Veen et
al. (\cite{vdVeen89}). We also constructed various color-color
diagrams using the near-infrared and \iras\ magnitudes. No distinction
can be made between the objects showing \brg\ in emission, absorption,
or a flat spectrum in the near and far-infrared color-color
diagrams. The near-infrared color-color diagrams show evidence for a
very large range of extinction, which in part is of circumstellar
origin.  Near-infrared versus far-infrared color-color diagrams show
trends that are consistent with the expected evolution of the
circumstellar shell.  This sample of post-AGB stars show a larger
range in color and are generally redder and closer to the galactic
plane than the ones known so far.

The properties of most of these objects are fully consistent with the
assumption that they are post-AGB stars that have not evolved far
enough yet to ionize a significant fraction of their circumstellar material.

\end{abstract}

\section{Introduction}

A most intriguing challenge is to understand how Asymptotic Giant
Branch (AGB) stars transform their surrounding mass-loss shells in a
couple of thousand years into the variety of shapes and sizes observed
in Planetary Nebulae (PNe).  There are a number of theories currently
being investigated. In the generalized interacting stellar wind 
model, a variety of axisymmetric PN shapes are obtained by the
interaction of a very fast central star wind with the progenitor AGB
circumstellar envelope (Kwok \cite{Kwok82}), when the latter is denser near the
equator than the poles (Frank et al. \cite{Frank93} ). Sahai \& Trauger (\cite{Sahai98})
proposed that the primary agent for shaping PNe is not the density
contrast, but a high speed collimated outflow of a few 100 km/s.
No consensus about the dominant physical process responsible for the
shaping of PNe has emerged so far, but there is agreement that they
occur during the early AGB-to-PN transition stage.  However, details of the
rate of evolution, the strength of the stellar wind, and the impact of
ionization across the transition phase are very poorly known.  In
order to test model predictions it is essential to study nebulae from
the early post-AGB phase through the very early PN phases.  However,
post-AGB objects are difficult to find, because this phase is very
short and the star is usually obscured by a thick
circumstellar dust shell.

One way of identifying new post-AGB stars is through their dust
emission.  In our search for new obscured PNe and post-AGB stars, we
selected candidates from the \iras\ Point Source Catalogue based on
infrared colors typical of PNe.  When we observed these candidates in
the radio continuum at 6~cm, on average 20~\% of the objects were
detected (Van de Steene \& Pottasch \cite{VdSteene93},
\cite{VdSteene95}).  Subsequent optical spectroscopy showed that the
PN candidates detected in the radio have emission line spectra typical
of PNe (Van de Steene et al.\ \cite{VdSteene96a}, \cite{VdSteene96b}).
However, the question remained: what is the evolutionary status of
those \iras\ sources with colors typical of PNe, which had no
detectable ionization in the radio (i.e. fluxes below 3~mJy)?  It is
possible that a few large, low surface-brightness PNe have escaped
detection.  Others of the remaining non-identified PN candidates could
be very young and small PNe. Their radio flux would have been below
our detection threshold and they wouldn't be identified in H$\alpha$
due to extinction.  However, the true evolutionary status of most
objects with \iras\ colors typical of PNe has remained unknown.

We have calculated evolutionary tracks for post-AGB stars in the
\iras\ color-color diagram (van Hoof et al.\ \cite{vHoof97}).  From
this study, it became clear that post-AGB objects and PNe could be
located in the same region in the \iras\ color-color diagram. No other
types of objects seem to have typically these particular \iras\ colors. We
therefore adopted the working hypothesis that the non-detected PN
candidates are AGB-to-PN transition objects. The goal of this
project is to determine the evolutionary status of these post-AGB candidates, 
their physical properties, and to relate them to their morphology.

To ensure the identification of the correct counterpart of the \iras\
source and obtain accurate positions for follow-up observations, we
first imaged the post-AGB candidates in the \jn-band. In order to
investigate whether the sources have some ionization, we obtained
\brg\ spectra. We reobserved the sources showing \brg\ emission
at 3~cm and 6~cm. 
To confirm the identification, obtain finding charts,
and photometry, we took high resolution near-infrared
(\jhkl) images.  

The next section describes the observations.  In Section~\ref{data} we
describe the data reduction and analysis, in Section~\ref{indiv} we
discuss the individual objects and in Section~\ref{discuss} we discuss
the general results.  We give a summary in the last section.

\section{Observations}
\label{observations}

\subsection{Imaging}

\subsubsection {\jn-band}

\begin{table}
\caption[]{Observation log for the \timmi\ observations.
The first column gives the object name, the second column the
observation date, the third column the total 
integration time in seconds, the fourth column gives the
\iras\ 12-$\mu$m band flux in jansky, and the last column
indicates whether the object was detected or not.}
\label{obs:log1}
\begin{tabular}{lrrll}
\hline
Name & Date & I.T.  & $S_{12}$ & detected \\
\iras       & dd/mm/yy & s  & Jy &  yes/no \\
\hline
10256$-$5628 & 10/05/95 & 155 & 4.30 & yes \\
11159$-$5954 & 10/05/95 & 466 & 1.08 & yes \\
11387$-$6113 & 11/05/95 & 404 & 3.56 & no \\
13356$-$6249 & 10/05/95 & 155 & 6.06 & yes \\
13416$-$6243 & 11/05/95 & 93 & 38.8 & yes \\
13428$-$6232 & 10/05/95 & 404 & 21.9 & yes \\
13428$-$6531 & 10/05/95 & 466 & 0.53 & no \\
13529$-$5934 & 10/05/95 & 466 & 1.37 & yes \\
14325$-$6428 & 10/05/95 & 218 & 3.31 & yes \\
14488$-$5405 & 11/05/95 & 155 & 6.59 & yes \\
15066$-$5532 & 10/05/95 & 466 & 1.43 & yes \\ 
15144$-$5812 & 11/05/95 & 218 & 7.67 & yes \\
15154$-$5258 & 11/05/95 & 684 & 2.25 & no \\
15310$-$6149 & 11/05/95 & 404 & 3.67 & no \\       
15544$-$5332 & 11/05/95 & 218 & 4.64 & yes \\
15553$-$5230 & 10/05/95 & 155 & 9.99 & yes \\ 
16086$-$5255 & 11/05/95 & 155 & 5.70 & yes \\
16115$-$5044 & 10/05/95 & 249 & 10.4 & no  \\
16127$-$5021 & 11/05/95 & 280 & 2.90 & yes \\
16130$-$4620 & 11/05/95 & 155 & 4.73 & yes \\
16209$-$4714 & 11/05/95 & 404 & 1.17 & no \\
16279$-$4757 & 10/05/95 & 93 & 43.0 & yes \\ 
16328$-$4517 & 11/05/95 & 280 & 3.73 & yes \\
16345$-$5001 & 11/05/95 & 218 & 2.45 & no \\
16594$-$4656 & 10/05/95 & 93 & 44.9 & yes \\ 
17009$-$4154 & 10/05/95 & 311 & 7.44 & yes \\
17088$-$4221 & 11/05/95 & 93 & 42.7 & yes \\
17234$-$4008 & 10/05/95 & 839 & 1.73 & no \\
\hline
\end{tabular}
\end{table}

Mid-infrared images in the \jn-band were obtained with the Thermal
Infrared Multi Mode Instrument (\timmi) at the European Southern
Observatory 10--12 May 1995.  The camera featured a 64~$\times$~64
element Gallium doped Silicon array with good cosmetic quality and a
quantum efficiency of 25~\%.  We used the \jn-band filter:
$\lambda_{\rm c}$~= 9.70~$\mu$m and $\Delta\lambda$~= 5.10~$\mu$m
with a setup yielding a pixel size of 0\farcs66, and a FOV of
$\approx$~40\arcsec.

For the cancelation of the strong background radiation, \timmi\ is
operated in chopping and nodding mode. The chopping frequency is
several hertz and nodding is done once or twice per minute.  In
chopping mode a pair of exposures is obtained: one exposure contains
object plus sky and the second exposure, 20\arcsec\ away, only sky.
To obtain the images in nodding mode the telescope was moved by an
angle exactly matching the chopper throw on the sky and the
observation in chopping mode was repeated.  Because this observation
is exactly 180 degrees out of phase with the first one, the observed
object shows up as a negative measurement.  Subtracting both images
resulted in a clean image corrected for first and second order effects
introduced by the strong and inhomogeneous background radiation.
To eliminate erratic pixels, the multiple
images of each object were median filtered with a high
threshold. Finally the images were Fourier transformed and the \midas\
standard low bandpass Butterworth filter was applied to better reveal
the sources.

A log of the observations can be found in Table~\ref{obs:log1}.  The
weather was poor during both nights. Consequently, only the brighter
\iras\ sources could be detected, while under favorable weather
conditions \timmi\ would be ten times more sensitive than \iras.  All
but one of the undetected sources have an \iras\ 12-$\mu$m flux below
4~Jy.  Six sources having a 12-$\mu$m flux below 4~Jy were detected
when the cloud conditions improved temporarily. The other 14 detected
sources have a larger \iras\ 12-$\mu$m flux.  We think that
non-detections are due to the cloudy weather rather than faulty \iras\
positions or inaccurate telescope pointing.


\subsubsection{Near-infrared images}

 \jhkl-band images of the 20 post-AGB candidates detected in the \jn-band
 were obtained with \caspir\ at the 2.3-m telescope at Siding
 Spring Observatory in Australia on 11 and
 12 March 1998. The detector is a SBRC CRC463 $256\times256$ InSb array
 which is sensitive from approximately 0.9~$\mu$m to 5.5~$\mu$m.  The
 pixel size was 0\farcs25 and the total field of view 1\arcmin.
 Each object was observed 60 times for 0.3~s at 
12\arcsec\ north and 12\arcsec\ south of its
 nominal position.  The objects were observed in
 order of right ascension: \iras\ 10256$-$5628 to 15144$-$5812
 on 11 March 1998 and \iras\ 15544$-$5332 to 17088$-$4221 on 12 March
 1998.  Both nights were photometric.  Standard stars were observed at
 the beginning of the observations each night. The seeing was
 sub-arcsecond.

\subsection{Observing the Br$\gamma$ line}

Using the improved positions from the \timmi\ images, we obtained
infrared spectra with \irspec\ at the
NTT at \eso\ 13 and 14 May 1995.

The spectra are centered at \brg\ and have a resolution of
$\lambda/\Delta\lambda \approx 2000$.  The SBRC $62\times58$ array
gives a total wavelength coverage from 2.15~$\mu$m to 2.18~$\mu$m.  The
resolution in the spatial direction is 2\farcs26/pixel.  Since the
outer edges of the array are vignetted, the total field of view is
approximately 1\farcm5. The slit width is 2 pixels or 4\farcs51.  The
slit orientation was north-south. First we pointed the telescope to a
bright star, and subsequently offset the telescope to the \timmi\
position.  The pointing accuracy of the NTT has an rms of about
1\arcsec.  Subsequently we moved the slit east- and westwards a few
arcseconds, to `peak-up' the strongest signal in order to center the
object in the slit. The resulting offsets were never larger than
6\arcsec.

To subtract the sky emission the beam-switching technique was used.
Because all but one (\iras\ 13428$-$6232) of the sources are much smaller
than 20\arcsec, two spectra were taken with the source at different
positions on the array, such that when the exposures were subtracted,
the spectra did not overlap.  For \iras\ 13428$-$6232 one integration was
taken on source and one on the sky and these images were subsequently
subtracted

The weather was very good, and the humidity was around 15~\%. Within
the accuracy limits of spectro-photometry the night was photometric,
even if the conditions slightly deteriorated at the end of the night.
Every 30 to 60~min a standard star was observed.
A complete log of the observations can be
found in Table~\ref{obs:log2}. Note that the observations include
four PNe from Van de Steene \& Pottasch (\cite{VdSteene95}).

\begin{table}
\caption[]{Observation log for the \irspec\ observations.
The first column gives the object name, the second column the
total on-source integration time in seconds, the third column the estimated
signal-to-noise ratio, the fourth column the airmass.}
\label{obs:log2}
\begin{tabular}{lrrrl}
\hline
Object & I.T. & S/N & Airm. & Comment \\
\hline
10256$-$5628 & 1200s &  170 & 1.135 & \\ 
11159$-$5954 &  600s &  970 & 1.172 & \\
13356$-$6249 & 1200s &  385 & 1.292 & \\ 
13416$-$6243 &  720s &  605 & 1.261 & \\ 
13428$-$6232 &  900s &  135 & 1.220 & \\ 
13529$-$5934 & 3240s &   32 & 1.166 & \\ 
14325$-$6428 &  600s &  165 & 1.226 & \\ 
14488$-$5405 &  720s &  225 & 1.104 & \\
15066$-$5532 &  720s &  130 & 1.127 & \\ 
15144$-$5812 &  900s &  340 & 1.163 & \\ 
15544$-$5332 &  720s &  370 & 1.106 & \\ 
15553$-$5230 & 1080s &   34 & 1.107 & \\ 
16086$-$5255 & 1080s &   56 & 1.118 & \\ 
16127$-$5021 & 1200s &  100 & 1.120 & \\ 
16130$-$4620 & 1080s &   50 & 1.134 & \\ 
16279$-$4757 &  360s & 1200 & 1.151 & \\ 
16328$-$4517 & 1080s &   80 & 1.150 & \\ 
16594$-$4656 & 1200s &  330 & 1.173 & \\ 
17009$-$4154 &  960s &  125 & 1.216 & \\ 
17088$-$4221 & 1080s &   45 & 1.263 & \\ 
\hline
18186$-$0833 &  600s &  235 & 1.375 & PN    \\
18231$-$1047 &  600s &  220 & 1.385 & PN    \\
18277$-$0729 & 1200s &  185 & 1.259 & PN    \\
18401$-$1109 &  480s &   50 & 1.372 & PN    \\
\hline
\end{tabular}
\end{table}

\subsection{Radio continuum observations}

The 6 objects showing \brg\ in emission were observed with the
Australia Telescope Compact Array for 12~h on 14 February 1997 and for
2~h on 15 February 1997.  The array configuration was $\#$6A
with the antennae at stations 3, 11, 16, 30, 34, and 37. The shortest
baseline was 337~m and the longest 5939~m.  The bandwidth was 128~MHz
divided in 32~channels centered at 4800~MHz and 8640~MHz corresponding
to 6~cm and 3~cm respectively. The sources were observed
simultaneously at both frequencies.  To obtain adequate spatial
coverage we cycled through the sample of 6 \iras\ objects and 3 calibrators
once an hour: 1338$-$59\,C, 13428$-$6232, 14488$-$5405, 1511$-$55\,C,
15144$-$5812, 15544$-$5332, 1600$-$48\,C, 16594$-$4656, 17009$-$4154
(with C denoting the calibrators).  Each object was observed for 8~min
and every phase calibrator for 3~min.  To avoid artifacts in the field
center we used an offset of 30\arcsec\ in declination.  At the
beginning and the end of each shift the primary flux density
calibrator 1934$-$638 was observed for 5~min to 10~min.  The weather was
unstable and humid.

\section{Data reduction and analysis} 
\label{data}

\subsection{Imaging}

\subsubsection{\jn-band}

At 10~$\mu$m, fields are not crowded: for each of our images we had
only one object within a relatively small field of view of 40\arcsec.
This leaves no doubt about the identification of the correct
counterpart in the \timmi\ field at this wavelength, even in
non-photometric weather conditions.  The disadvantage is that there
are no other points of reference to improve upon the \iras\
coordinates other than the telescope pointing.

The positions and Full Width at Half Maximum (FWHM) of the sources
were determined from the images after filtering.  Because of the
variable sky conditions, the background level around the sources was
badly defined. Consequently the determination of the (standard) star's
PSF was very uncertain. To remedy this, we subtracted the background
in the following way: we made an image in which we replaced a circular area
surrounding the source by a flat surface fitted to the background
outside this circle and subtracted this image from the original. 
The resulting frame contained only the object against a virtual zero background. 
Next a two dimensional Gaussian was fitted to the source
to determine the position and FWHM more accurately.

At regular intervals during the two nights, we pointed the telescope
to a standard star and put it exactly on the cross hairs. Next, we
took an image to determine exactly the corresponding position on the
array.  Similarly, for each object, we calibrated the telescope
pointing using the closest bright SAO star, before doing an offset to
the \iras\ position.  There was always only one source in the field,
which left no doubt about the identification of the \iras\ source in
the \jn-band image.  We assumed that the \iras\ position corresponds to
the reference position on the array as determined from the standard stars.  
We measured the offset from this array position to the position of the
source in the \timmi\ image, and adopted this as the improved position
of the \iras\ source.  The difference between the \iras\ and \timmi\
positions were mostly less than 10\arcsec\ which is in agreement with
what was found in positional difference between the radio detections
and the \iras\ positions (Van de Steene \& Pottasch
\cite{VdSteene93}, \cite{VdSteene95}). 
The uncertainty in the \timmi\ positions was estimated to be less than
$\sim$5\arcsec\ and is due to the fact that in these regions with high
extinction, the nearest SAO star could be several arcminutes away and
the pointing of the \eso\ 3.6-m was not very good.

The median seeing was 2\farcs7 FWHM during the first night and, even
worse, 3\farcs4 FWHM, during the second. Fifteen objects remained
unresolved.  Five appear to be extended at 10~$\mu$m.  The contour
plots of the extended sources are presented in Fig.~\ref{plotsN}.  The
morphology of the resolved objects is similar to young PNe.  The size
of the extended sources is given in Table~\ref{tabmorph}.  One showed
\brg\ in absorption, two in emission, and in two we didn't detect
any \brg\ in absorption or emission.

\subsubsection{Near-infrared}

The \jhkl-band images were reduced in \iraf\ using standard procedures
as described in the \caspir\ manual (McGregor \cite{McGregor94}).  The
two images taken 24\arcsec\ apart in declination were
combined. Because the objects were most prominent in the \jk-band, the
resulting \jk-band images are presented in Fig.~\ref{Kimages}.

The positions of the objects determined from these images are listed
in Table~\ref{pos-size}.  We used the {\sc imcoor} package in \iraf\
and the positions of reference stars from the United States Naval
Observatory Catalogue (\usno\ release A1.0, available via the \eso\
{\sc skycat} tool) to determine an accurate position.  If the object
had a counterpart in the catalogue, we adopted the
\usno\ catalogue position. The positional uncertainty is similar to
the uncertainty in the \usno\ positions, about 1\arcsec. The object
for which we have obtained the \brg\ spectrum is indicated with a
box in Fig.~\ref{Kimages}. Note that in some fields more than one star 
was in the slit. The \iras\ position is indicated with a cross 
in Fig.~\ref{Kimages}. 

Although the K-band seeing was sub-arcsecond, most of the objects are
unresolved at this level. The sizes of the extended objects are
presented in Table~\ref{tabmorph}. All objects which are extended in the
\jk-band show \brg\ in emission, but not vice versa.  
Obviously, comparison of morphology in the \jn- and \jk-band 
is not a good identification tool.

The photometry was done with the program {\sc qphot} in \iraf.
The \jhkl-band magnitudes of the objects were determined for each of
the two images separately, and their agreement checked before
averaging. The average difference between the two measurements was
0.01~mag for both nights in \jhk\ and 0.05~mag for the \jl-band. The
near-infrared magnitudes are presented in Table~\ref{flux:tab}. The
uncertainties are estimated at 0.05~mag in \jhk\ and 0.1~mag in the
\jl-band, including measurements errors and the uncertainty in the
correction for atmospheric extinction.  Table~\ref{flux:tab} also
contains estimates for the \jk\ magnitudes derived from the \brg\
spectra (see also Sect.~\ref{sec:brg}).
Usually these magnitudes are a bit fainter
than the ones determined from the \caspir\ images. This is probably 
due to slit loss. 

Table~\ref{flux:tab} also presents near-infrared photometry from the
literature, when available. The literature values are based on
aperture photometry rather than imaging.  In crowded regions they are
often brighter, likely due to contamination by neighboring objects.
In all cases where the literature value was very different from our magnitude, the two
values were not associated with the same star, and we were able to
identify the star measured by the other authors in our images.

 Near-infrared imaging is preferred to aperture photometry to identify
 the object in the field. It can provide very accurate coordinates and
 finding charts for follow-up observations. Given photometric weather and good
 seeing conditions, the magnitudes won't be contaminated by neighboring
 stars. When several near-infrared bands are available, colors can be
 used as a secondary means of identification.  The counterpart often
 is the reddest, but not necessarily the brightest object in the near
 infrared.  Due to their thick circumstellar dust shells, post-AGB
 stars often are very reddened.

\begin{table}
\caption{The angular sizes of the 6 extended sources.
The first column gives the name of the object, the second the 
 extent in arcsecond in RA and DEC in the \jn-band image, 
the third similarly the visible extent in the \jk-band image, and
the last whether \brg\ was seen in absorption (A), emission (E), or 
the object had a flat spectrum at \brg\ (F). Unresolved objects
are indicated by PS. }
\begin{flushleft}
\begin{tabular}{lrrr}
\hline
Name    & \jn-band & \jk$-$ band & \brg \\
        & arcsec$^2$ & arcsec$^2$ & \\
\hline
13356$-$6249 & 5.1 $\times$ 6.0 & PS &  A \\ 
13428$-$6232 & 4.1 $\times$ 12.0 & 6.0 $\times$ 11.0 & E \\
15066$-$5532 & 5.9 $\times$ 6.5 & PS & F \\
15553$-$5230 & PS & 3.0 $\times$ 1.5 & E \\
16086$-$5255 & 6.3 $\times$ 9.5 & PS & F \\
17009$-$4154 & 8.4 $\times$ 7.0 & 6.8 $\times$ 6.8 & E \\
\hline
\end{tabular}
\end{flushleft}
\label{tabmorph}
\end{table}

\begin{table*}
\caption[]{ Coordinate list of the \iras\ sources. If the object was
associated with a star in a catalogue, the catalogue position is given.
Otherwise the position as determined from our imaging is given. The abbreviations
in front of the catalogue numbers have the following meaning: \usnos\ -- \usnof,
\gscs\ -- \gscf, \deniss\ -- \denisf.
The galactic longitude and latitude in degrees are given in columns 2 \& 3 respectively.}
\label{pos-size}
\begin{tabular}{lrrrrrrl}
\hline
Name & \ctr{1}{$l^{\rm II}$}{-2} & \ctr{1}{$b^{\rm II}$}{-4} & \multicolumn{2}{c}{Position (J2000)} & \multicolumn{2}{c}{Catalogue Position (J2000)} & comment \\ 
     & degrees & degrees  & RA(h m s) & DEC($^\circ$ \arcmin\ \arcsec) & RA(h m s) & DEC($^\circ$ \arcmin\ \arcsec) & \\ 
\hline
10256$-$5628     &  284.1410  & 0.7907  &  &  & 10 27 35.23 & $-$56 44 19.7 & \usnos\ 0300$-$09688714 \\
11159$-$5954     &   291.5727 & 0.6201  &  &  & 11 18 07.12 & $-$60 10 38.5 & \usnos\ 0225$-$10943841 \\
13356$-$6249     & 308.2971 & $-$0.7047 & 13 39 06.39 & $-$63 04 43.8 & 13 39 06.38 & $-$63 04 43.5 & \deniss\ J133906.4$-$630443 \\
13416$-$6243     & 308.9871 & $-$0.7324 & 13 45 07.28 & $-$62 58 16.7 & & & \\
13428$-$6232     & 309.1598 & $-$0.5935 &  13 46 20.94 & $-$62 47 57.7 & & & \\
13529$-$5934$^a$ & 311.0217 &  2.0315 &  13 56 24.62 & $-$59 48 57.0 & & & North \\
                 & 311.0218 & 2.0305 &  13 56 24.78 & $-$59 49 00.5 & & & South  \\ 
14325$-$6428     &     313.8718 & $-$4.0772 &  &  & 14 36 34.42 & $-$64 41 31.4 & \usnos\ 0225$-$20526456 \\
                 &   &  &     &               & 14 36 34.50 & $-$64 41 30.7 & \deniss\ J143634.5$-$644131 \\
                 &   &  &     &               & 14 36 34.39 & $-$64 41 31.2 & \gscs\ 0901500077 \\
14488$-$5405     &   320.0898   & 4.4941 &       &               & 14 52 28.75 & $-$54 17 43.0 & \usnos\ 0300$-$22393022 \\
                 &        &    &   &               & 14 52 28.73 & $-$54 17 43.2 & \gscs\ 0868000930 \\
15066$-$5532$^a$ & 321.6609 & 1.9965 & 15 10 26.08 & $-$55 44 13.9 & & & West \\
                 &   321.6623 & 1.9962  &        &               & 15 10 26.65 & $-$55 44 12.2 & East, \usnos\ 0300$-$23036148 \\     
15144$-$5812     & 321.2041  &  $-$0.8267 &15 18 21.84 & $-$58 23 11.8 & & & \\
15544$-$5332     & 328.4769  &  $-$0.3422 &15 58 18.75 & $-$53 40 39.9 & & & \\
15553$-$5230     & 329.2468  &   0.3602 & 15 59 10.70 & $-$52 38 37.2 & & & \\
16086$-$5255     &  330.4722 &  $-$1.2876   &        &       & 16 12 30.47 & $-$53 03 09.2 & \usnos\ 0300$-$27060539 \\
16127$-$5021     & 332.6920  &  0.1477 & 16 16 30.27 & $-$50 28 57.3 & & & \\
16130$-$4620$^a$ & 335.5096  & 3.0125 & 16 16 42.85 & $-$46 27 55.5 & & & North \\
                 & 335.5096  &  3.0114 & 16 16 42.96 & $-$46 28 00.1 & & & South  \\
16279$-$4757     & 336.1443   & 0.0833 & 16 31 38.76 & $-$48 04 05.7 & & & \\
16328$-$4517     & 338.6604 &   1.2902 & 16 36 25.75 & $-$45 24 03.1 & & & \\
16594$-$4656     &   340.3924  & $-$3.2889   &       &               & 17 03 10.03 & $-$47 00 27.8 & \usnos\ 0375$-$29889766 \\
                 &       &  &       &               & 17 03 10.00 & $-$47 00 26.9 & Hrivnak et al.\ (\cite{Hrivnak99}) \\
17009$-$4154     & 344.5342 &  $-$0.4193 & 17 04 29.59 & $-$41 58 35.9 & & & \\
17088$-$4221$^a$ &   345.0504  & $-$1.8521     &     &               & 17 12 21.69 & $-$42 25 09.0 & North, \usnos\ 0450$-$26723221\\
                 &    345.0471 &  $-$1.8572    &  &               & 17 12 22.38 & $-$42 25 29.4 & South, \usnos\ 0450$-$26723710 \\
\hline
\end{tabular}
\ntd{$^a$The correct identification is uncertain; two possible counterparts are given.}
\end{table*}

\subsection{The Br$\gamma$ spectra}
\label{sec:brg}

\begin{table*}
\caption[]{Johnson \jhkl\ magnitudes for the program stars.
Columns~2, 3, 5 and 8 give the \jhkl\ magnitudes determined from the \caspir\ images.
Column~4 gives the \jk-band magnitude estimated from the continuum 
flux at 2.166~$\mu$m measured by \irspec.
Column~6 gives literature
values for the \jk\ magnitude, if available, and column~7 the references.
Column~9 gives the SED class defined by van der Veen et al.\ (\protect\cite{vdVeen89}).}
\label{flux:tab}
\begin{tabular}{lrr|rrrl|rrl}
\hline
Name & \jj & \jh & \jk$_{\rm irspec}$ & \jk$_{\rm caspir}$ & \jk$_{\rm lit}$ & ref & \jl & SED  & Comment  \\
     & mag & mag & mag              & mag              &  mag          &     & mag & \class\    &       \\
\hline
10256$-$5628 &  10.93 & 9.81 & 9.16 &  9.05 &             9.14 &       1  &  8.29 & IVa & \\
11159$-$5954 &   9.45 & 7.99 & 6.85 &  7.15 &                  &          &  6.16 & IVa & variable ? \\
13356$-$6249 &  9.38  & 7.83 &  7.20 &  6.97 &       6.97, 6.95 &    1, 4 &   6.2 & IVa & \\
13416$-$6243 &  10.17 & 8.59 &  7.64 &  7.43 & 7.64, 7.58, 7.52 & 3, 2, 2 & 5.37 & II & variable \\
13428$-$6232 &  12.50 & 10.35 & $<$9.52 &  8.78 & 9.08, 9.41, 9.07 & 1, 1, 3 &    7.40 & IVa$^\prime$ & \\
13529$-$5934$^a$ & 13.85 & 12.65 & --- &  12.23     &                  &         &  10.59 &  & North \\
                 & 14.50 & 13.09 & --- &  12.22     &                  &         &   --- &  & South \\
14325$-$6428 & 9.27 & 8.81 &   8.78 &  8.61 &             8.64 &       4 &   8.27 & IVb & \\
14488$-$5405 & 8.89 & 8.53 &   8.57 &  8.30 &                  &         &   7.80 & IVb & \\ 
15066$-$5532$^a$ & 11.23 & 9.61 &   9.11 &  8.95 &   8.64 &       1 &  8.35 &  & West \\
                 &  10.89 & 10.19 & ---  &   9.46   &    &   &   9.37 &  & East \\
15144$-$5812 &  11.00 & 8.95 &  8.20 &  7.20 &                  &         &   5.08 & II & variable ?  \\
15544$-$5332 &  14.00 & 10.43 &  8.24 &  7.90 &                  &         &   5.30 & II & \\
15553$-$5230 &  13.99 & 11.58 & 10.57 & 10.03 &                  &         &   8.01 & II & \\
16086$-$5255 &  10.68 & 9.91 &  9.76 &  9.73 &           7.25       &  2    &  9.52 & IVa/b & \\
16127$-$5021 &  12.83 & 10.92 & 8.80 &  8.57 &                  &          &  6.76 & II & \\
16130$-$4620$^a$ &  14.80 & 12.76 & --- &  11.06  &                &         &   8.76 & & North \\
                 & 11.73 & 10.37 & --- &  9.95 &             9.63 &       2 &  9.48 & & South \\
16279$-$4757 & 8.84 & 6.83 & $<$6.59 &  5.74 &       5.63, 5.62 &    2, 3 &  4.51 & IVa$^\prime$ &  variable ?\\ 
16328$-$4517 & 11.43 & 10.58 &  10.10 & 10.02 &                  &         &   9.17 & IVa & \\
16594$-$4656 & 9.73 & 8.85 & $<$8.62 &  8.20 &       8.21, 8.17 &    1, 3 &   6.91 & IVa & \\
17009$-$4154 & 12.75 & 10.44 & $<$9.27 &  9.00 &       8.75, 9.28 &       1 &   6.99 & II & variable \\
17088$-$4221$^a$ & 10.90 & 9.54  & --- &  9.08  &       9.09, 9.22 &    1, 3 &   8.96   &  &  North \\
                 & 11.25 & 10.35 & 10.04   &  10.07  &        &    &  9.68 &  & South \\
\hline
\end{tabular}
\ntd{$^a$The correct identification is uncertain; two possible counterparts are given.}
\ntd{References ---
1.~Garc\'{\i}a-Lario et al.\ (\cite{Garcia97}),
2.~Hu et al.\ (\cite{Hu93}),
3.~van der Veen et al.\ (\cite{vdVeen89}),
4.~\denis\ project (Epchtein et al. \cite{Epch94}).
}
\end{table*}

The data reduction was done using the standard reduction macros
contained in the \midas\ image processing system.  After flatfielding,
sky subtraction, and rectifying the spectrum the resultant image
contained a positive and a negative spectrum of the source. The next
steps were to extract the positive and negative beam, invert the
negative beam, and average the two.  The spectra were accurately
wavelength calibrated using the sky-emission lines present in the
non-subtracted images. The rest wavelength of these lines were taken
from Oliva \& Origlia (\cite{Oliva92}). Using the standard star spectra,
a flux conversion factor was determined to calibrate all spectra.

The values of the continuum flux at \brg\ (2.166~$\mu$m)
have been converted to Johnson \jk-band magnitudes for all sources
and are listed in Table~\ref{flux:tab}. Since the central wavelength
of the \jk-band filter (2.20~$\mu$m) nearly coincides with the central
wavelength of our spectra, no attempt has been made to correct for the
different slopes in the continuum. The error introduced by this
assumption is well within the accuracy of the flux calibration.
The \jk-band fluxes of four sources have been marked as a lower limit.
For \iras\ 16279$-$4757 and \iras\ 16594$-$4656 the reason is that the
absorption profile extends beyond the observed spectral range, hence
the estimated continuum is a lower limit to the true level. \iras\
13428$-$6232 and \iras\ 17009$-$4154 are larger than the slit at these
wavelengths. Flux will have been missed for other sources as well,
because the peak-up procedure to position the object in the slit was
not very accurate.  No attempt has been made to correct for slit loss,
because it was impossible to optically verify how well the object was
centered in the slit.

When a clear continuum was present, the spectra were
normalized by fitting a second order polynomial to the continuum and
dividing the spectrum by the fit. The spectra of \iras\ 16279$-$4757
and \iras\ 16594$-$4656 showed a very wide \brg\ absorption and no or
hardly any continuum. Therefore, the spectrum of \iras\ 16279$-$4757
was divided by the maximum flux present in the spectrum.
For \iras\ 16594$-$4656 some continuum
seemed present and a linear fit was made between both ends of the
spectrum. This normalization should be considered uncertain.
The continuum of the four detected PNe is very weak and
therefore they were not normalized.

A Voigt profile was used to fit the \brg\ absorption lines in the
normalized spectrum. From the fits, the equivalent widths $W_{\lambda}$
were determined using: \\
\[ W_{\lambda} = \int_{0}^{\infty} [1 - r(\lambda)] \, {\rm d}\lambda, \] 
where $r(\lambda)$ is the residual flux normalized to 1 at the continuum. The
emission lines were unresolved at our instrumental resolution, and could be
fitted well with a Gaussian profile. Table~\ref{hydr:fit} lists the equivalent
widths of the absorption and emission profiles, along with the Doppler FWHM
$\Delta\nu_{\rm D}$, and the Lorentz FWHM $\Delta\nu_{\rm L}$~= $\Gamma/2\pi$
(where $\Gamma$ is the effective damping constant) of the Voigt profile.


For \iras\ 14488$-$5405 and \iras\ 16594$-$4656, both absorption and emission
components were present. A combination of a Voigt and a Gaussian profile, each
with its own central wavelength, was used in the fit for these objects. For
\iras\ 16594$-$4656, the central part of the absorption and the emission were
fitted well, but the outer wings of the absorption were not. For \iras\
16086$-$5255, a weak absorption line at the central wavelength of \brg\ was
observed.  It could not fitted well.  For \iras\ 11159$-$5954, and \iras\
15553$-$5230, no \brg\ was seen in emission or absorption. For \iras\
16279$-$4757 the resulting `best fit' was so bad that no parameters are
listed. The fact that the absorption lines in \iras\ 16279$-$4757 and \iras\
16594$-$4656 cannot be fitted properly with a Voigt profile, indicates the
presence of an additional broadening mechanism. One likely candidate is the
linear Stark effect.
Further investigations are needed to verify this, but if
confirmed, this would indicate a high surface gravity of the central
star.

For the spectra containing emission features, the absolute line fluxes
were determined. In most spectra only \brg\ was detected, but some
also showed the presence of He\,{\sc i} emission. In these spectra the
the He\,{\sc i} $4f-7g$ $^1\!F^\circ-^1\!G$ and $^3\!F^\circ-^3\!G$ 
multiplets would be blended with the \brg\ line, while the He\,{\sc i}
$4d-7f$ $^1\!D-^1\!F^\circ$ and $^3\!D-^3\!F^\circ$ blend could be
seen separately.  It is unlikely that He\,{\sc ii} $8-14$ line was
also blended with \brg, except maybe for high excitation PNe ($T_{\rm
eff} > 80\,000$~K), such as \iras\ 18401$-$1109. 
 In the fitting procedure, it was assumed that each line had a 
Gaussian profile with the same (unresolved) width 
and that the wavelength interval between the lines was fixed. The continuum was
assumed to be flat except for \iras\ 14488$-$5405 and \iras\
16594$-$4656 where the underlying \brg\ absorption was assumed to have
a Voigt profile. The helium lines were only fitted when there were
indications for their presence upon visual inspection.  The separate
flux values in the decomposition of the \brg\ complex could not be
determined accurately and therefore only the total flux of the blend
is listed.  The He\,{\sc i} lines never contributed more than 8.5~\%
to the total flux of this blend.  The results of these fits are listed in Table~\ref{abs:emm}. 
 No attempt has been made to correct any of the fluxes
for slit loss. 

The central wavelengths of the hydrogen absorption and emission
features can be used to calculate the heliocentric velocities of these
objects, using the routine {\sc rvcorrect} in \iraf. The results can
be found in Table~\ref{hydr:fit}. The literature values for the radial
velocities of the standard stars (Hirshfeld et al. \cite{Hirsh91}) were
compared with their observed radial velocities, yielding an average
accuracy of 11~\kms.

\begin{table*}
\caption[]{Parameters for the fits to the \brg\ lines. The 2nd and 3rd columns
list the central wavelength and the equivalent width of the absorption
feature. The 4th and 5th column give the same information for the emission
feature. The 6th and 7th column give the Doppler and Lorentz FWHM of the
Voigt profile (accuracy $\sim$5~\kms, except entries with a colon). The
8th column gives the total flux of the H\,{\sc i} $4-7$, He\,{\sc i} $4f-7g$, He\,{\sc ii} $8-14$
blend (the helium lines only being included where indicated), and the 9th column the
flux of the He\,{\sc i} $4d-7f$ 
blend. Fluxes of sources larger than the slit are marked as lower limits.
The last column gives the heliocentric radial velocities (accuracy $\sim$10~\kms).
If two entries are given, they are the radial velocities from the absorption
and emission line, respectively.}
\label{hydr:fit}
\label{abs:emm}
\begin{tabular}{lrr@{\hspace{6mm}}rr@{\hspace{4mm}}rr@{\hspace{2mm}}r@{\hspace{4mm}}rr}
\hline
Name & \ctr{1}{$\lambda_{\rm abs}$}{0} & \ctr{1}{$W_{\lambda}^{\rm abs}$}{0} & \ctr{1}{$\lambda_{\rm em}$}{0} & \ctr{1}{$W_{\lambda}^{\rm em}$}{0} & \ctr{1}{$\Delta\nu_{\rm D}$}{-2} & \ctr{1}{$\Delta\nu_{\rm L}$}{-1} & \ctr{1}{H\,{\sc i}}{0} & \ctr{1}{He\,{\sc i}}{-1} & $v_{\rm rad}$ \\
\iras\ & \ctr{1}{$\mu$m}{0} & \ctr{1}{\AA}{0} & \ctr{1}{$\mu$m}{0} & \ctr{1}{\AA}{0} & 
\ctr{1}{\kms}{0} & \ctr{1}{\kms}{0} & \ctr{2}{$10^{-18}$~W\,m$^{-2}$}{0} & \kms \\
\hline
10256$-$5628 & 2.16635 &    4.80    &         &           &   33 &  166   &           &         &   104 \\
13356$-$6249 & 2.16534 &    5.89    &         &           &   25 &  196   &           &         & $-$26 \\
13416$-$6243 & 2.16518 &    3.34    &         &           &   60 &  183   &           &         & $-$48 \\
13428$-$6232 &         &            & 2.16549 &  $-$6.17  &      &        &   $>39.0$ &         &  $-$5 \\
14325$-$6428 & 2.16499 &    5.44    &         &           &   16 &  208   &           &         & $-$72 \\
14488$-$5405 & 2.16583 &    4.47\h  & 2.16574 &  $-$2.82  &   85 &  281   &      43.0 &         & 44, 32\\
15144$-$5812 &         &            & 2.16589 &  $-$8.93  &      &        &    190.\f &         &    55 \\
15544$-$5332 &         &            & 2.16499 &  $-$1.21  &      &        &      24.9 &         & $-$68 \\
16127$-$5021 & 2.16559 &    2.76    &         &           &  121 &   73   &           &         &    17 \\
16328$-$4517 & 2.16477 &    5.47    &         &           &   30 &  202   &           &         & $-$95 \\
16594$-$4656 & 2.16511 &   15.6\h\f & 2.16563 & $-$5.23\h &  0\h & 1107\h &      76.6 &         &$-$46, 26\\
17009$-$4154 &         &            & 2.16518 & $-$17.9\f &      &        &  170.$^a$ &  11.3\h & $-$36 \\
\hline
18186$-$0833 &         &            & 2.16470 &           &      &        &  526.$^a$ &  14.4   & $-$96 \\ 
18231$-$1047 &         &            & 2.16572 &           &      &        &  427.$^a$ &  13.9   &    46 \\ 
18277$-$0729 &         &            & 2.16641 &           &      &        &  261.$^a$ &   3.0\h &   142 \\ 
18401$-$1109 &         &            & 2.16555 &           &      &        & $>129.^a$ & $>5.5$\h&    24 \\ 
\hline
\end{tabular}
\ntd{$^a$ Components for the He\,{\sc i} lines were included in the fit. }
\end{table*}


\subsection{Radio observations}
\label{radiobs}

The data were reduced using the package {\sc miriad} following
standard reduction steps as described in the reference guide by Bob
Sault and Neil Killeen
(http://www.atnf.csiro.au/Software/Analysis/miriad).  
Images were made using the multi-frequency synthesis technique and
robust weighting with a robustness parameter of 0.5. 
Any confusing sources were {\sc clean}-ed before determining the upper
limits and noise in the map.  

We didn't detect any of the six emission line sources above a
detection limit of 0.70~mJy/beam at 3~cm and 0.55~mJy/beam at 6~cm.
Our individual upper limits to the flux for the sources are given in
Table~\ref{radio}.  
The weakest confusing source which was clearly detected was 1~mJy and
the strongest 3.8~mJy.  The map of \iras\ 15544$-$5332 showed large
scale structure, but still no source was present after deleting the
two shortest baselines. 

Using the \brg\ flux and assuming Case~B recombination we
calculated the expected optically thin radio flux.  These values are
given in the last column of Table~\ref{radio}.  The predicted values
appear to be a factor ten or more higher than the upper limits on the
observed flux. Even the flux per beam of the two extended objects
would have been well above the detection limit at 6 cm. For the PNe,
however, the predicted values are somewhat lower than what was
observed in the radio.  The \brg\ flux is probably underestimated
due to extinction and slit loss. The radio flux has an uncertainty of
10~\% to 20~\%.

\begin{table}
\caption[]{Results of the radio continuum observations.
The total on source integration time was 96~min.
Column 2 and 3 give the upper limit and
rms at 6~cm, column 4 and 5 the upper limit and rms at 3~cm,
and the last column the radio flux predicted on basis of the observed 
\brg\ flux, assuming Case~B recombination. The last four objects are
known PNe. For these objects only the observed and predicted 6~cm
flux are given.}
\label{radio}
\begin{tabular}{lrrrrr}
\hline
Name & \multicolumn{2}{l}{S$_{6cm}$} & \multicolumn{2}{l}{S$_{3cm}$} & S$_{6cm}$  \\
     & \multicolumn{2}{l}{mJy/beam} & \multicolumn{2}{l}{mJy/beam} & mJy \\
           & U.L. & rms  & U.L. & rms & predicted \\
\hline
13428$-$6232   & 0.43   & 0.11  &  0.42 & 0.12  & $>$4.8 \\
14488$-$5405   & 0.42  & 0.12  &  0.58 & 0.15 & 5.3 \\
15144$-$5812   & 0.53  & 0.12   & 0.66 & 0.15 & 23.6 \\
15544$-$5332   & 0.46  & 0.13   & 0.51 & 0.14 & 3.1 \\
16594$-$4656   & 0.53   & 0.13   & 0.75 & 0.18 & 9.5 \\
17009$-$4154   & 0.55   & 0.13    & 0.67 & 0.16 & 19.3 \\
\hline
Name & \multicolumn{2}{l}{S$_{6cm}$ (mJy)} & \multicolumn{2}{l}{} & S$_{6cm}$ (mJy) \\
             & \multicolumn{2}{l}{observed}  &  &  & predicted \\
\hline
18186$-$0833 & 87.2 & & & & 62.5 \\ 
18231$-$1047 & 90.5 & & & & 49.9 \\
18277$-$0729 & 44.1 & & & & 31.7 \\
18401$-$1109 & 36.\f & & & & $>$15.3 \\
\hline
\end{tabular}
\end{table}

\section{Discussion of individual \iras\ objects}
\label{indiv}

{\bf 10256$-$5628:} The position of the \iras\ counterpart agrees
with that of a m(red)~= 17.1~mag star in the \usno\ catalogue.  The
\jk\ magnitude of Garc\'{\i}a-Lario et al.\ (\cite{Garcia97}) is in
agreement with our value.
 
\noindent
{\bf 11159$-$5954:}
This object has an \usno\ counterpart with m(red)~= 21.0~mag.
This is the only star for which the \jk\ magnitude derived
from the \brg\ spectrum is significantly brighter than
the one determined from the \jk-band image. Hence this star may be
variable. Its optical spectrum shows that this is an M-type star (Van de Steene
et al., in preparation).

\noindent
{\bf 13356$-$6249:} Our value for the \jk\ magnitude is in perfect
agreement with the magnitude mentioned by Garc\'{\i}a-Lario et
al.\ (\cite{Garcia97}).  The position and magnitude are also in
excellent agreement with the data published by the \denis\
project (Epchtein et al. \cite{Epch94}). 
This object appears extended in our \timmi\ image.

\noindent
{\bf 13416$-$6342:} 
According to Hu et al.\ (\cite{Hu93}), the source is a highly reddened 
G1I star with \jr~= 17.4~mag. Taking into account the photometry of Hu et al.\ (\cite{Hu93}),
this object seems to have become brighter 
since the 1987 observations of van der Veen et al.\ (\cite{vdVeen89})
by $\Delta$\jj~= 0.74~mag, $\Delta$\jh~= 0.46~mag  and $\Delta$\jk~= 0.21~mag respectively.
However in the \jl-band the star became fainter by 0.30~mag.

\noindent
{\bf 13428$-$6232:} This is a very nice bipolar showing \brg\ in
emission. The magnitude of van der Veen et al.\ (\cite{vdVeen89}) is somewhat
fainter than ours. Considering its size, possibly part of the flux was
missed in their 10\arcsec\ diaphragm.

\noindent
{\bf 13529$-$5934:}
In the \caspir\ images we see two faint stars less than 4\arcsec\
apart.  The top one is closest to the \iras\ position and the reddest
source, and therefore the most likely counterpart. The correct
identification of the \iras\ counterpart needs to be confirmed at
10~$\mu$m.  The spectra of the objects are completely blended in the
\irspec\ image.  Their combined spectrum shows no \brg\ in
absorption or emission.  We won't discuss this spectrum any further.

\noindent
{\bf 14325$-$6428:}
This object is associated with a bright \gsc\ star of \jv~= 11.8~mag. 
It was also observed by the \denis\ project. 
Their \jk$_{\rm s}$ magnitude is in good agreement with ours.

\noindent
{\bf 14488$-$5405:} This \iras\ object is also associated with
a bright \jv~= 11.5~mag \gsc\ star. This star shows \brg\ in emission.

\noindent
{\bf 15066$-$5532:}
The counterpart at the \iras\ position is a bright m(red)~= 14~mag \usno\
star. Its magnitude is \jk~= 9.46~mag. However, our \timmi\
position, and hence the object for which we obtained a spectrum with \irspec, 
is 5\farcs5 to the west.  It is the reddest of the two and its 
magnitude is \jk~= 8.95~mag. It showed a flat \brg\ spectrum. The
object appeared extended in our \timmi\ image.  The \jk\ magnitude
determined by Garc\'{\i}a-Lario et al.\ (\cite{Garcia97}) is brighter
than each one of ours, but he may have had both neighbors in its
15\arcsec\ aperture. The correct identification of the \iras\
counterpart needs to be confirmed at 10~$\mu$m. We won't discuss this
spectrum any further.

\noindent
{\bf 15144$-$5812:} 
The \jk\ magnitude determined in the \caspir\ image is 1~mag brighter
than the magnitude determined from the \irspec\ continuum spectrum. We
see this bright object and the fainter star towards the south both in
the slit. Because of this and the \brg\ in emission, easily
visible in the spectrum of the brightest star, there is no doubt that
the difference in magnitude is not due to an identification
error. Hence this source could be variable.
 

\noindent
{\bf 15553$-$5230:} Our \jk\ magnitude is 4~mag fainter than the value
determined by Garc\'{\i}a-Lario et al.\ (\cite{Garcia97}). 
These determinations cannot correspond to the same source.  In the
\jk-band image we see four objects close together.  The \jk\ magnitude of the
brightest star to the west is \jk~= 6.21~mag and does agree with the value
determined by Garc\'{\i}a-Lario et al.\ (\cite{Garcia97}).  There is a
fainter, very red, extended object just eastward of this bright
source.  It has an elliptical morphology.  
Most likely this is the correct \iras\ counterpart.

\noindent
{\bf 16086$-$5255:} Our \usno\ counterpart is significantly fainter
than the magnitude \jk~= 7.30~mag determined by Hu et
al. (\cite{Hu93}).  Their magnitude is in agreement with 
our magnitude of the bright, red  \usno\ star to the south. Its spectral
type would be M3I. Our position is within 3\arcsec\ of the \iras\
position, while the southern star is more than
16\arcsec\ away. We therefore adopt our
identification as the true \iras\ counterpart.



\noindent
{\bf 16130$-$4620:} In the \jk-band image the object corresponds to two
sources 4\arcsec\ apart. Hu et al.\ (\cite{Hu93}) associated the
southern star (\jv~= 16.7~mag) with an M5Ib star. Our magnitude is
in agreement with his measurement.
However the top one is by far the reddest of the two, invisible in the
optical. This makes it a much stronger candidate for being the \iras\
counterpart.  The spectra of both objects were blended into one
extended source in the \irspec\ observations. We shall not discuss
this spectrum any further.

\noindent
{\bf 16279$-$4757:} Our \jk\ magnitude is 0.1~mag fainter and our
\jl\ magnitude 0.7~mag fainter than the values measured by van der Veen et
al. in 1987 and Hu et al. in 1990. However, the \jj- and \jh-band values are
in perfect agreement.  This difference could be due to measurement
errors, or the object may be variable.  Hu et al.\
(\cite{Hu93}) determined \jr~= 18.4~mag and classified it as a G5 star
based on its optical spectrum.


\noindent
{\bf 16594$-$4656:} Our magnitude is in perfect agreement with what
was measured by Garc\'{\i}a-Lario et al.\ (\cite{Garcia97}).  \hst\
images show the presence of a bright central star surrounded by a
multiple-axis bipolar nebulosity with a complex morphology and a size
of 6\farcs3 $\times$ 3\farcs3 (Hrivnak et al.\
\cite{Hrivnak99}). In our \jk-band images we do not see any evidence of 
the bipolar nebula. The counterpart of this object is a m(red)~= 12.4~mag 
\usno\ star.

\noindent
{\bf 17009$-$4154:} Our magnitudes obtained in 1998 are fainter
than the 1990 value of Garc\'{\i}a-Lario et al.\ (\cite{Garcia97}), but
brighter than their 1992 values. This source seems variable. It is a
post-AGB star with \brg\ in emission surrounded by a faint
nebulosity. The object is 7\arcsec\ in diameter and would have fitted in the aperture 
of Garc\'{\i}a-Lario, if well centered.

\noindent
{\bf 17088$-$4221:}  This source has been observed by van der Veen et al.\
(\cite{vdVeen89}) and Garc\'{\i}a-Lario et al.\ (\cite{Garcia97}).
Their \jk\ magnitudes are about 1~mag brighter than ours. The position given
by van der Veen et al.\ (\cite{vdVeen89}) corresponds to
the bright, red \usno\ star west of the \iras\ position. 
We measured a magnitude \jk~= 9.08~mag for this star in our image, in
agreement with their value. The star for which we obtained the
\brg\ spectrum corresponds to the \usno\ star south of the \iras\ 
position. This object is further away from the \iras\ position.
The northern star is more likely to be the \iras\ counterpart, although
the correct identification needs to be confirmed 
at 10~$\mu$m. We won't discuss this object any further.


\section{Discussion}
\label{discuss}

\subsection{Identification of the \iras\ counterpart}

Due to poor telescope pointing of the \eso\ \mbox{3.6-m} at the time
of the observations, the counterparts of 3 \iras\ sources
(13529$-$5934, 15066$-$5532, and 17088$-$4221) could not be
determined unambiguously, even after careful analysis of the
images. Due to the poor spatial resolution of
\irspec, the spectrum of \iras\ 16130$-$4620 was useless due to blending.
We won't consider these objects any further, so that
16 objects remain for further discussion.

\subsection{Br$\gamma$}

Based on the \brg\ spectra we can identify objects of three
different types: those with a \brg\ emission line, those with a
\brg\ absorption line, and those with no \brg\ line at all.
We detected \brg\ in absorption in 7 out of 16 objects.  The
absorption lines are very narrow in 6 objects, indicating a low
surface gravity. This is a strong indication for the post-AGB nature
of these objects.  Six objects show \brg\ in emission. Two of
these also show a photospheric absorption profile.  All emission line
sources have a strong underlying continuum, unlike normal PNe.  In
another three objects no clear \brg\ absorption or emission was
visible.

\subsection{Radio continuum}

As noted in Sect.~\ref{radiobs}, the predicted optically-thin radio
flux values (assuming Case~B recombination) appear to be at least a
factor of ten higher than the observational flux limits for the
post-AGB objects.  This indicates that for these objects either the
radio flux is optically thick at 3~cm, or the Case~B assumption is not
valid.

After a star leaves the AGB, its mass loss decreases by several orders
of magnitude while simultaneously the velocity of the wind increases.
Hence the star will be surrounded by an increasingly more tenuous wind
inside a detached AGB shell. At first the star will be too cool to
cause any significant ionization, either in the wind or in the AGB
shell.  However, as the stellar temperature increases, ionization will
start and the ionization front will steadily move outwards. This will
also be the case for the radius at which the 3~cm radiation becomes
optically thick (for brevity we will call this the 3~cm
radius, $r_{\rm 3cm}$). Optically thick radio emission from a sphere
of radius  $r_{\rm 3cm}$ and a temperature T$_e$ gives a flux density
at frequency $\nu$ of 

\[ S_{\nu} = \frac{r_{\rm 3cm}^2}{D^2} \frac{2\pi\nu^2kT_{\rm e}}{c^2}. \]
where D is the distance to the star.

Using this approximation
we calculated that $r_{\rm 3cm} <$ 3$\times10^{14}$~cm,
assuming $S_{\rm 3cm} <$ 0.7~mJy, $T_{\rm e}$~= $10^4$~K, and $D$~= 1~kpc.
This upper limit is similar to the size Knapp et al.\
(\cite{Knapp95}) determined for the post-AGB stars CRL~915 ($S_{\rm 3cm}=$0.30~mJy)
and \iras\ 17423$-$1755 ($S_{\rm 3cm}=$0.44~mJy)  
using the same assumptions.  As long as the
ionization front has not reached the AGB shell, the 3~cm radius will
not change much, because the outer regions in the $1/r^2$ density
profile contribute very little to the optical depth, and
consequently the radio flux will remain very low. However, once the
post-AGB star reaches a temperature hot enough to ionize the AGB
shell, the 3~cm radius will quickly increase by roughly two orders of
magnitude, causing a dramatic increase in radio flux. This marks the
onset of the PN phase. As an example, the sizes of the young
PNe CRL~618 and \iras\ 21282+5050 measured by 
Knapp et al.\ (\cite{Knapp95}) are a
factor 10 larger than the sizes of post-AGB star nebulae. Their radio
flux values, 67~mJy and 4.3~mJy respectively, are well above our
detection limit.

In our post-AGB star candidates with \brg\ in emission, the ionized
region where the emission originates must be very small and dense.
Probably, the AGB shell of our objects is not yet ionized, but the
post-AGB wind could be. The \brg\ spectra are unresolved at a
resolution of $\sim$150~\kms. Hence the wind velocity couldn't be much
larger than this value. Evidence for a wind emanating from some of these
central stars was presented in Van de Steene et al. (\cite{VdSteene00}).

\subsection{Spectral energy distribution}

 One of the well-established characteristics of post-AGB stars is that
 their Spectral Energy Distributions (SEDs) have a `double-peaked'
 shape.  The two peaks in the
 spectrum correspond to the stellar and dust emission
 components. Post-AGB stars have been classified into four
 classes based on the shape of the SED by van der Veen et
 al.\ (\cite{vdVeen89}). 

\begin{itemize}
\item \class~I: has a flat spectrum between 4~$\mu$m and 25~$\mu$m and a steep fall-off
to shorter wavelengths.
\item \class~II: maximum around 25~$\mu$m and a gradual fall-off to shorter 
wavelengths.
\item \class~III: maximum around 25~$\mu$m and a steep fall-off to a 
plateau roughly between 1~$\mu$m and 4~$\mu$m with a steep fall-off at shorter
wavelengths.
\item \class~IV: two distinct maxima, one around 25~$\mu$m and a second between
2~$\mu$m and 3~$\mu$m (IVa$^\prime$), between 1~$\mu$m and 2~$\mu$m (IVa), 
or below 1~$\mu$m (IVb).
\end{itemize}

Note that \class~IVa$^\prime$ was not contained in the original
definition, but was added to classify objects that did not fit in any
of the original categories.

The SEDs of the objects are shown in Fig.~10. The objects have the
typical post-AGB SEDs as cited above. The SED class of each object is
listed in Table~\ref{flux:tab}.  Six of the 16 positively identified
objects are of \class~II and the other 10 of \class~IV.  For objects
in \class~II the circumstellar dust is so optically thick that almost
all star light is absorbed by the dust and is re-radiated at mid- to
far-infrared wavelengths. The large infrared excess is commonly
attributed to the presence of a very compact circumstellar dust shell
and/or ongoing mass loss which obscures the central star from view.
Objects in \class~IV have less obscured central stars: the thermal
emission from their circumstellar shells appear as a peak in the
far-infrared and the central stars show up as a peak in the
near-infrared (IVa) or optical (IVb).  We always see some stellar
signature in the near-infrared and therefore have no objects of
\class~III. For instance, 
Van der Veen et al. (\cite{vdVeen89}) classified \iras\
16594-4656 as \class~III, while we classified it as IVa.  
We extended the definition of \class~IVa to include objects
which show a clear stellar signature in the near-infrared, but 
peak in the \jk-band, just beyond 2~$\mu$m. 
(e.g. \iras\ 13428-6232, \iras\ 16279-4757). In the table
these objects are marked as \class~IVa$^\prime$. These objects 
have no optical counterpart in the \usno\ catalog.
When objects in \class~IV peak at shorter wavelengths they usually have an
\usno\ counterpart.  The central stars of objects in \class~IVb are
less reddened and brighter than objects in \class~IVa and 
have bright optical counterparts (e.g.  \iras\ 14325-6428, \iras\ 14488$-$5405).

We especially draw attention to two objects which have unique SEDs.
\iras\ 15544$-$5332 is the only object in the sample for which 
 the \jl-band value is higher than the \iras\ 12~$\mu$m value. It has the
coolest dust shell in the sample. It also has 
a very steep \class~II spectrum, indicative of a very high extinction.
\iras\ 11159-5954 is the only object in the sample
for which the peak of its SED in the near-infrared is higher than in
the far infrared, showing that the grain emission is very weak.

\subsection{Color-color diagrams}

\subsubsection{\iras\ color-color diagram}

\begin{figure}
\caption{\iras\ color-color diagram.  The squares represent objects
having \brg\ in absorption, the asterisks objects having \brg\ in emission
and the crosses flat spectrum sources. Diamonds indicate sources for which the
near-infrared counterpart is not certain. The objects are labeled by
the first four numbers of their \iras\ name. The SED class of each object
is indicated next to its symbol. The boxes as defined by van der 
Veen \& Habing (\cite{vdVeen88}) are drawn in.}
\center{\epsfxsize=9cm \epsfbox{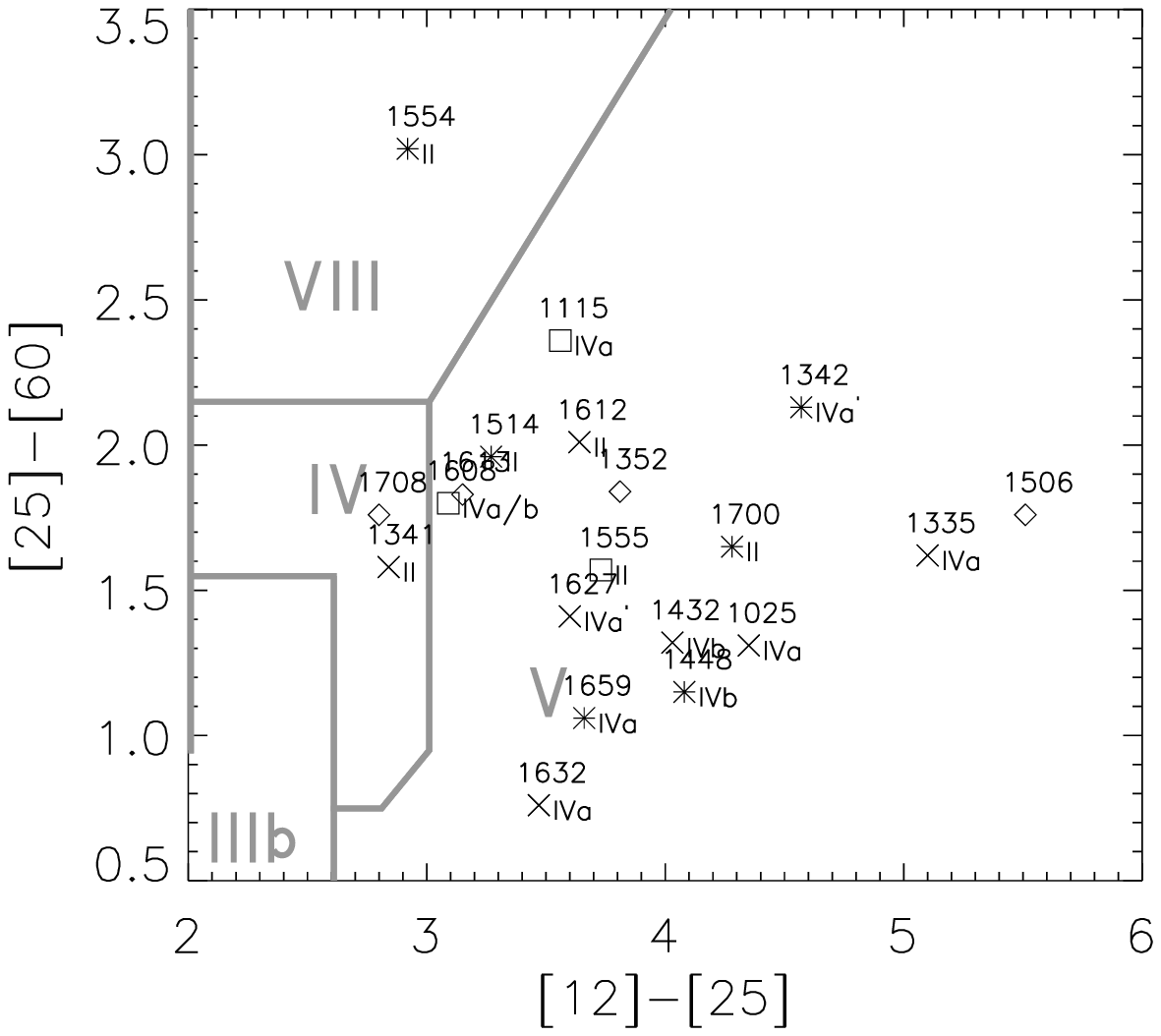}}
\label{figfir}
\end{figure}

In Fig.~\ref{figfir} we show the \iras\ color-color diagram.  The
\iras\ fluxes were converted to magnitudes according to the \iras\ 
Explanatory Supplement (Beichman et al.\ \cite{Beichman84}). 
The boxes defined by van der Veen \& Habing
(\cite{vdVeen88}) are drawn in. According to this classification scheme,
 PNe are found
in region~V of the color-color diagram and AGB stars in region~IV.
In region~VIII there may be some confusion from galaxies and young
stellar objects, and in region~IV an odd H\,{\sc ii} region may be
present.  The only object in region~VIII is \iras\ 15544$-$5332, which is not
redshifted and shows \brg\ in emission. The object is unresolved,
and the \brg\ emission is very weak. Hence it is unlikely to be an
ultra-compact H\,{\sc ii} region, but we cannot completely rule out
that it is an embedded young stellar object that is not hot enough
to ionize its surroundings. 
\iras\ 13416$-$6243 in region~IV 
has \brg\ in absorption and hence can be
considered to be a post-AGB star. 
>From the results in Fig.~\ref{figfir}, it seems that 
in the \iras\ color-color diagram no distinction can be made between
post-AGB stars with \brg\ in emission, absorption, or a flat
\brg\ spectrum. Post-AGB stars were expected to be
located in a region in the \iras\ color-color diagram between AGB
stars and PNe (e.g. Volk \& Kwok \cite{Volk89}; van der
Veen et al. \cite{vdVeen89}; Hu et al. \cite{Hu93}).  However van Hoof et al.\
(\cite{vHoof97}) found in their parameter study of the spectral
evolution of post-AGB stars that they can follow a variety
of paths in the \iras\ color-color diagram. Consequently PNe and
post-AGB stars can occupy the same region in the \iras\ color-color
diagram and the position in the \iras\ color-color diagram alone
cannot {\it a priori} give a unique determination of the evolutionary status
of a post-AGB star. Our observations confirm this result. 

The fact that our objects are mostly selected from the region where
PNe were found and not in the region between AGB and PNe, may explain
our high detection rate of \brg\ emission (for comparison, the
search by K\"aufl et al. (\cite{Kaufl93}) for Br$\alpha$ emission in a
sample of 21 post-AGB stars resulted in only one detection).

\subsubsection{Near-infrared color-color diagrams}

\begin{figure}
\caption{\jj$-$\jh\ versus \jh$-$\jk\ diagram. 
The symbols are defined in Fig.~\ref{figfir}. The arrow shows
the effect of correcting for $A_V$~= 5~mag.}
\center{\epsfxsize=9cm \epsfbox{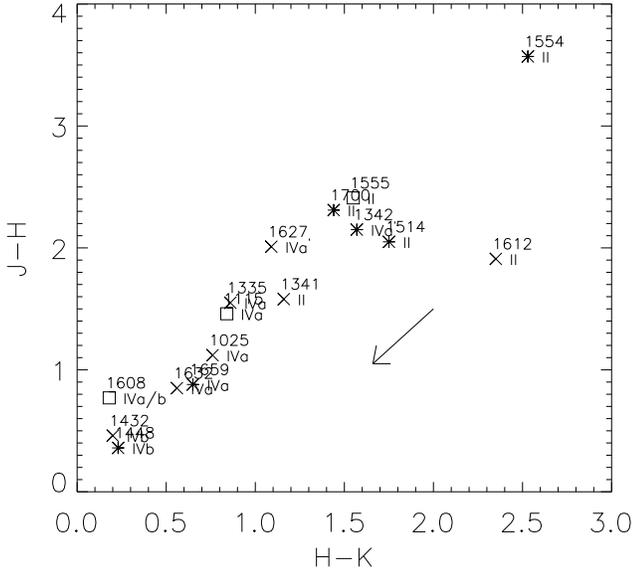}}
\label{fignir1}
\end{figure}

\begin{figure}
\caption{\jh$-$\jk\ versus \jk$-$\jl\ diagram. 
The symbols are defined in Figs.~\ref{figfir} and \ref{fignir1}.}
\center{\epsfxsize=9cm \epsfbox{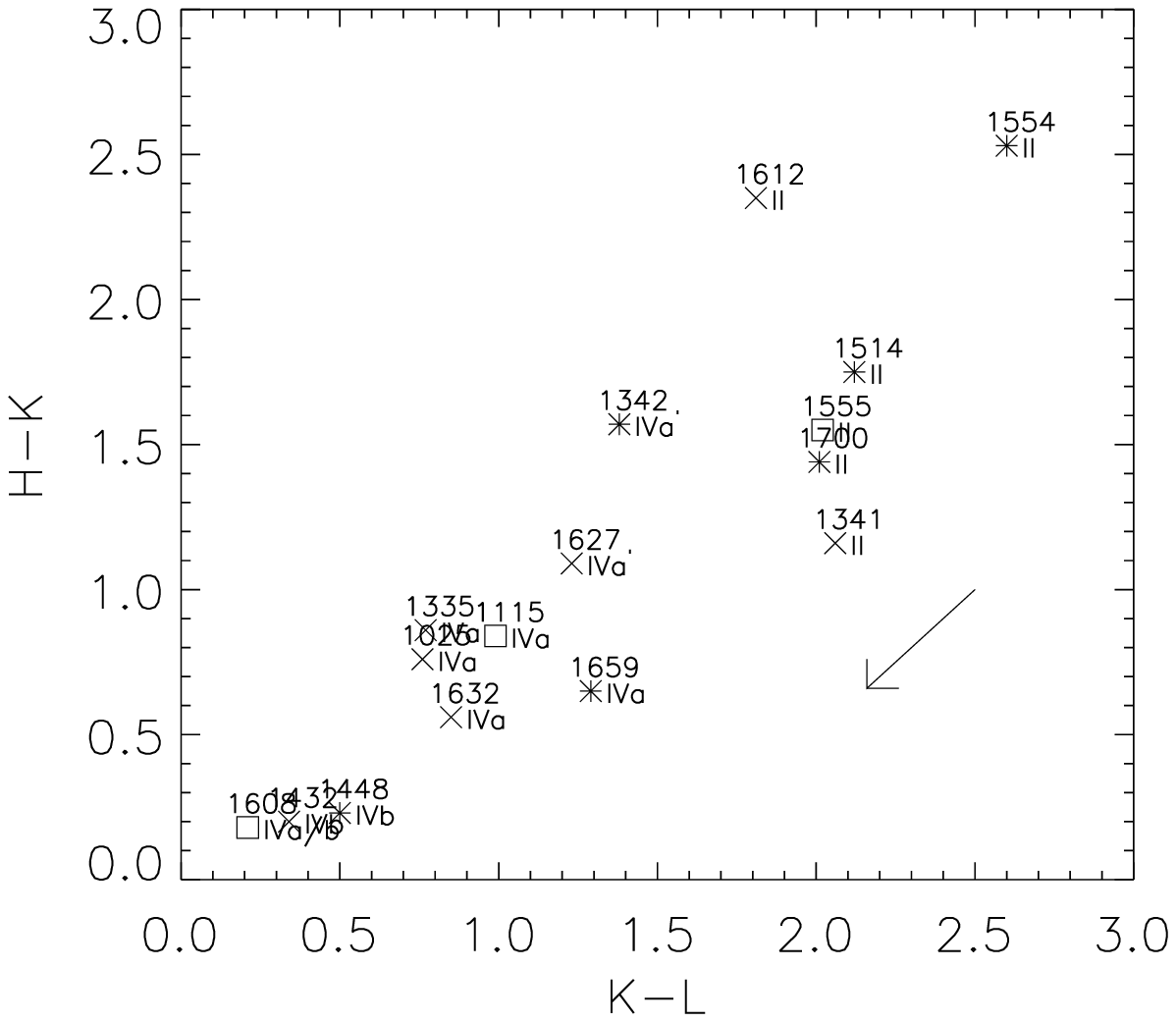}}
\label{fignir2}
\end{figure}

In Fig.~\ref{fignir1} and Fig.~\ref{fignir2} we show the \jj$-$\jh\
versus \jh$-$\jk\ and the \jh$-$\jk\ versus \jk$-$\jl\ color-color
diagrams, respectively. In the diagrams we notice that
objects of \class~II are redder than the objects of \class~IV. The
former are found in the top right part of the diagrams while the
latter are located more towards the bottom left. The separation is
most obvious in Fig.~\ref{fignir2}.  For our sample, objects having
\jk$-$\jl~$>$ 1.5~mag are all of \class~II and objects which
have a maximum in the near-infrared beyond 2~$\mu$m have 
\jh$-$\jk~$>$ 1.0~mag. Objects for which the
stellar signature is more pronounced and which peak more towards shorter
wavelengths in the near-infrared, are found more towards the bottom
left in the diagrams.

To understand this effect, we would like to point out that for all but
the coolest stars the intrinsic shape of the stellar continuum in the
near-infrared can be approximated by a Rayleigh-Jeans tail. Since the
shape of this tail does not depend on stellar temperature, the
intrinsic infrared colors of these stars will not depend on stellar
temperature either. Furthermore, since the colors of an A0V star are
by definition zero, the infrared colors for most other stars are also
close to zero, except when severe inter- or circum-stellar extinction
is present. In this case the infrared colors will be positive and can
be used as a crude measure for the extinction.

The mass loss rate during the superwind phase is very high. At the end
of this phase the circumstellar dust will almost completely obscure the
central star. During the post-AGB evolution, as the AGB shell expands,
the shell will become more optically thin to stellar radiation. The
stellar signature becomes more pronounced and will peak towards ever
shorter wavelengths in the near-infrared.
Consequently, it is expected that post-AGB stars will move from the
upper right of the diagram to the lower left as the AGB shell expands
and the circumstellar extinction becomes less.  Generally speaking, we
might expect that the objects in the top right of the diagram have
left the AGB more recently than the objects in the lower left.
However, we need to be cautious about such an interpretation because
of the combined effects of many unknowns.  The time it takes for the
envelope to become optically thin will depend upon the mass loss rate
at the tip of the AGB, the wind velocity, and the distribution of the
mass in the circumstellar shell.  If the AGB star had a non-spherical
mass loss concentrated towards the equator, the central star
could be visible along polar directions, while being
completely obscured in equatorial directions. Hence, the observed amount of
circumstellar extinction will depend on the
viewing angle (Soker \cite{Soker99}). The interstellar
extinction also affects the position of the objects in the
color-color diagrams, as indicated by the arrow.
The \jj$-$\jh\ versus \jh$-$\jk\ diagram is clearly
the most affected by interstellar extinction.
Eleven of the 16 objects are within one degree of the galactic plane,
including the 6 objects in \class~II. The two objects in \class~IVb 
also have the highest latitude ($|b| >$ $4^\circ$). We calculated the extinction
for our objects at a distance of 1~kpc and 4~kpc according to Hakkila et al.
(\cite{Hakkila97}): $A_V$ would be between 0.5~mag and 2.0~mag if the objects were at a distance of 1~kpc and 
between 2.1~mag and 6.2~mag with a median of 4.6~mag if they were at 4~kpc.


Objects with \brg\ in emission, absorption, or a flat spectrum
are well mixed in the diagrams. As discussed in the previous section
it is unlikely that the stars with \brg\ in emission have reached
a temperature high enough to start to ionize the AGB shell and the
emission probably originates in the stellar wind.  Because we observe
\brg\ in emission from objects in \class~II (e.g. \iras\
15144$-$5812), it seems that a fast stellar wind can be present at an
early stage when the circumstellar shell is still very optically
thick.  We are currently trying to determine spectral types of our
objects in order to determine their true post-AGB evolutionary status
(Van de Steene et al., in preparation).

\subsubsection{Combined near- and far-infrared color-color diagrams}

\begin{figure}
\caption{\jk$-$\jl\ versus [12]$-$[25] diagram. 
The symbols are defined in Figs.~\ref{figfir} and \ref{fignir1}.}
\center{\epsfxsize=9cm \epsfbox{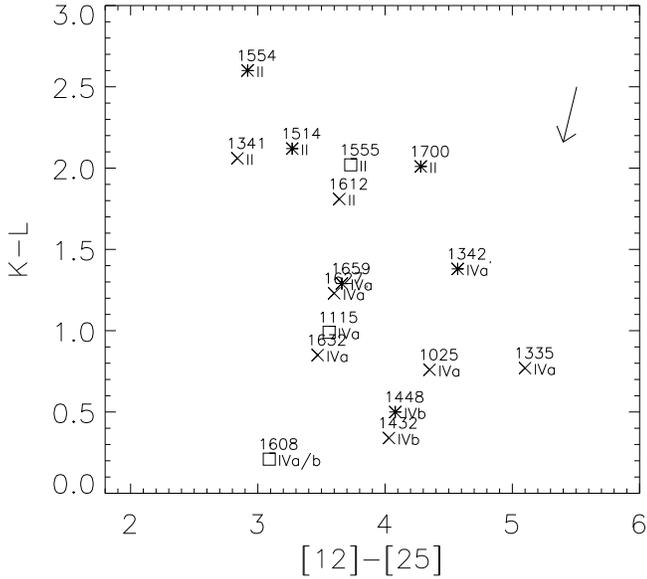}}
\label{figstd1}
\end{figure}

\begin{figure}
\caption{\jk$-$\jl\ versus [25]$-$[60] diagram. 
The symbols are defined in Figs.~\ref{figfir} and \ref{fignir1}.}
\center{\epsfxsize=9cm \epsfbox{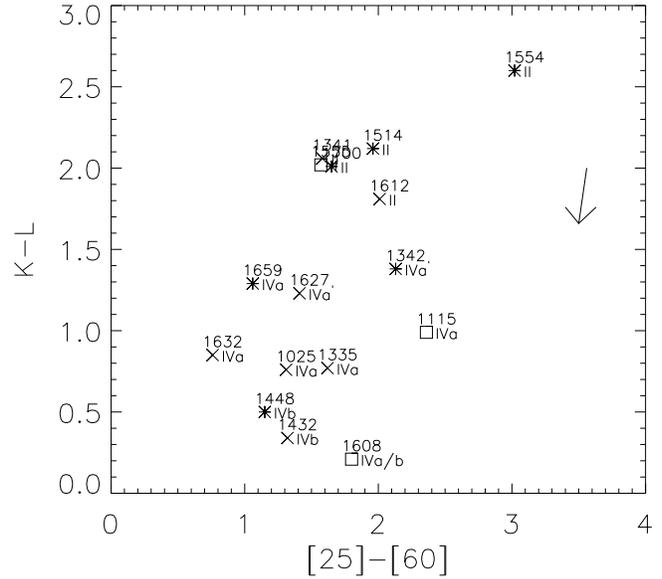}}
\label{figstd2}
\end{figure}

In Fig.~\ref{figstd1} and Fig.~\ref{figstd2} we show the \jk$-$\jl\
versus [12]$-$[25] and \jk$-$\jl\ versus [25]$-$[60] diagrams,
respectively.  The \jk$-$\jl\ color roughly describes the evolution of
the circumstellar extinction and is therefore a measure for the
expansion of the AGB shell, while the [12]$-$[25] and [25]$-$[60] colors
reflect the evolution of the spectrum of the circumstellar dust shell.

In Fig.~\ref{figstd1} we see a weak trend towards cooler [12]$-$[25]
colors with decreasing \jk$-$\jl\ color at first. When
\jk$-$\jl~$<$ 1.5~mag this trend seems to reverse,
probably due to an increase in 12~$\mu$m flux. However,
this trend will need to be confirmed with a larger sample.

In Fig.~\ref{figstd2} we see a broad and weak trend towards hotter
[25]$-$[60] colors with decreasing \jk$-$\jl\ color. Obviously the
25-$\mu$m flux increases faster than the 60-$\mu$m flux.

Theoretical calculations by Bl\"ocker (\cite{Bloecker95}) predict that
directly after the star leaves the AGB, the stellar evolution is slow.
This causes the AGB dust shell to cool when it expands. However, around
$T_{\rm eff} \approx 8000$~K the evolution of the central star speeds up
considerably and the grains in the AGB shell start to heat up again.
This effect is most pronounced for silicate grains because, as the peak
of the stellar spectrum moves into the UV, the efficiency with which
these grains absorb light increases significantly. This causes the
counter-clockwise loop which was first predicted by van Hoof et al.
(\cite{vHoof97}) for oxygen-rich post-AGB stars in the \iras\
color-color diagram. This effect could
explain the reverse trend in Fig.~\ref{figstd1}, and also the trend in
Fig.~\ref{figstd2}.

\begin{figure}
\caption{\jk$-$\jl\ versus \jl$-$[25] color-color diagram. 
The symbols are defined in Figs.~\ref{figfir} and \ref{fignir1}. The dotted and dashed
lines are explained in the text.}
\center{\epsfxsize=9cm \epsfbox{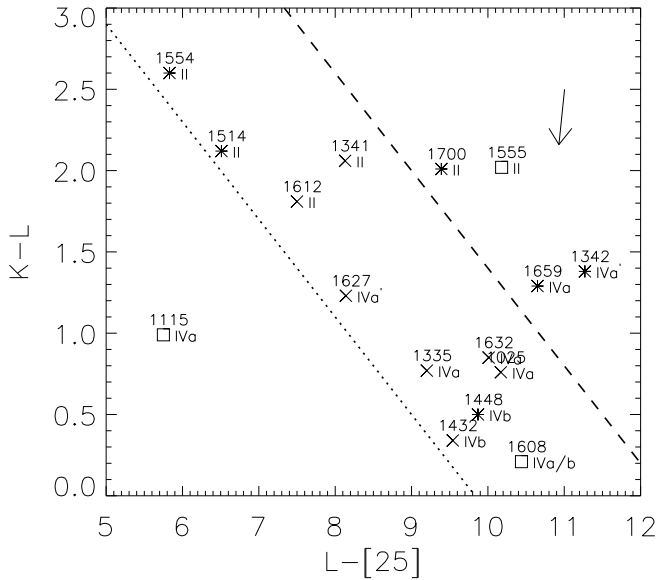}}
\label{figkl25}
\end{figure}

In Fig.~\ref{figkl25} we plot the \jk$-$\jl\ versus \jl$-$[25]
diagram.  The \jk$-$\jl\ color describes the evolution of the
circumstellar extinction.  The \jl$-$[25] color relates the stellar
component with the peak of the dust emission.

Because he \jk$-$\jl\ color is sensitive to the extinction and the
\jl$-$[25] color is insensitive to extinction, all \class~II objects
are found in the upper half of the diagram and \class~IV objects in
the lower half.

For our limited sample, all but one of the objects are found to the
right of the dotted line. This is partly due to sample selection.
As the arrow indicates, objects less obscured  (e.g., 
due to orientation along the polar axis, or because of an optically thin shell, 
or less interstellar extinction),
and especially the ones with cool central stars, 
could be situated below the dotted line.

\iras\ 11159$-$5954 has a lower extinction than its \jl$-$[25] color
would indicate.  This is an M-type star (Van de Steene et al., 2000,
in preparation), which may still have ongoing mass loss. Its dust shell
 does not appear to be very thick and the star is very bright in the near
infrared. 

The objects to the right of the dashed line
are extended in the near-infrared or the optical.
\iras\ 17009$-$4154 and \iras\ 15553$-$5230 showed elliptical morphology.
The former is very faint in the optical, the latter invisible.
\iras\ 13428$-$6232 shows a bipolar morphology in the near-infrared and
and is also very faint in the optical. \iras\ 16594$-$4656 is bright
and was not observed to be extended in the near-infrared, 
though it showed a bipolar morphology in its \hst\ image 
(Hrivnak et al. \cite{Hrivnak99}). Probably they have a thicker cool
circumstellar dust shell than objects located below the dashed line (more
25~$\mu$m-band flux) and/or their central star temperatures are higher 
(less \jl-band flux).

As the dust shell expands, the [25] magnitude will decrease a bit (for
oxygen-rich post-AGB stars) or remain roughly constant (for
carbon-rich post-AGB stars) (van Hoof et al. \cite{vHoof97}).  The
\jl\ magnitude will increase, as the stellar temperature increases.
The \jk\ magnitude will decrease as the star starts shining through
the dust shell.  Thus the \jl$-$[25] values are expected to increase
with decreasing \jk$-$\jl\ values as the shell expands and the star
becomes hotter.

This is what we observe in Fig.~\ref{figkl25}, both for the extended
and non-extended objects.  A small \jl$-$[25] color indicates that the
system is young, hence the star is heavily obscured by the dust shell
which is still close to the star. This is noticeable by the large
\jk$-$\jl\ values. For objects with a larger \jl$-$[25] color the AGB
shell is already more detached and cooler.  The extinction is less and
this translates into a smaller value for the \jk$-$\jl\
color. 

The \jk$-$\jl\ versus \jl$-$[12] diagram (not shown) is very similar to 
the \jk$-$\jl\ versus \jl$-$[25] diagram, but 
the spread is larger for \jk$-$\jl~$<$ 1.0~mag, possibly because of
an increase in 12-$\mu$m flux, as discussed in the previous section.  
Because of its relatively 
large 12-$\mu$m flux, \iras\ 13416$-$6342  is located
close to \iras\ 17009$-$4154. It needs to be checked at
higher resolution whether this source is extended.

\begin{figure}
\caption{\jj$-$\jk\ versus \jk$-$[25] color-color diagram. 
The symbols are defined in Figs.~\ref{figfir} and \ref{fignir1}. The grey regions
show the location of objects in Ueta et al.(\cite{Ueta00}).
These regions are labeled according to their definitions
as {\it SOLE} and {\it DUPLEX} with and without star.
The dashed line separates the two kind of objects.}
\center{\epsfxsize=9cm \epsfbox{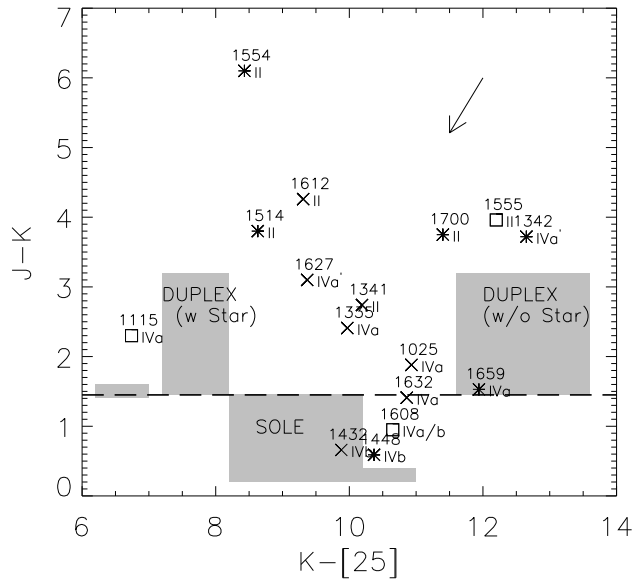}}
\label{figjk25}
\end{figure}

In Fig.~\ref{figjk25} we show the color-color diagram proposed by Ueta
et al.  (\cite{Ueta00}). Since the \jj$-$\jk\ color measures the
circumstellar extinction, in a similar way as \jk$-$\jl\ color does,
and because the evolution described above for the
\jl$-$[25] color is equally valid for the \jk$-$[25] color, one would expect
the same decreasing trend as in the previous diagram.  However, 
extinction effects have a much stronger effect on
the \jk$-$[25] and \jj$-$\jk\ colors than on the \jl$-$[25] and
\jk$-$\jl\ colors, as can be seen from the arrows in both diagrams.
Nearly all objects of \class~IVa have \jj$-$\jk~$<$ 2.5~mag.
 When the star becomes prominent in the near-infrared, 
the stellar maximum shifts through the \jk-band towards the \jj-band,
causing a reversal in the evolutionary trend. \\
The extended objects are also found towards the right of the diagram.
In this plot it is more obvious that \iras\ 16594$-$4656 is the
brightest of the 4 extended objects.

Ueta et al. (\cite{Ueta00}) had one M~star, \iras\ 04386$+$5722,
offset towards blue \jk$-$[25] color, similar to \iras\ 11159$-$5954,
but a bit bluer in \jj$-$\jk. Its position is indicated with the
smallest grey box in Fig.~\ref{figjk25}.

The long-dashed line at \jj$-$\jk~= 1.45~mag in Fig.~\ref{figjk25}
separates the regions what Ueta et al. call {\it Star-Obvious Low
level Elongated} ({\it SOLE}) and {\it DUst Prominent Longitudinally
EXtended} ({\it DUPLEX}) nebulae. One quarter of our objects would be
classified as {\it SOLE} and three quarters as {\it DUPLEX} in this scheme.  The
grey regions show where the objects in their sample are located and
are labeled with their acronyms.  Our samples are obviously
complementary: there is virtually no overlap !  The difference may be
caused in part by the selection criteria.  They have selected known
post-AGB candidates from the literature and imaged their nebulosities
with WFPC2 in the optical.  Consequently they chose optically bright
post-AGB stars.
  We selected objects from the PNe region in the \iras\
color-color diagram of which very few had an optical counterpart
identified.  Moreover, all our objects are located within 5 degrees of
the galactic plane, while only 11 out of their 27 objects are (3 {\it
SOLE}, 6 {\it DUPLEX}, and 2 stellar).  The arrow in the diagram
indicates a correction for $A_V$~= 5~mag of extinction. Larger
interstellar extinction alone cannot explain why our samples
appear different. 
If the objects in both samples are at similar distances, it is plausible 
that we have more massive central stars in our sample, and that they are evolving faster across
the HR diagram. 
Further investigation is needed to understand the differences between both
samples.

In summary, no distinction can be made between the objects showing
\brg\ in emission, absorption, or a flat spectrum, in any of the
color-color diagrams.  The trends we see in the near and far infrared
are mainly due to the expansion, morphology, and dust properties in the
circumstellar shell and the obscuration of the central star it causes.
The trends show the expected evolution of the circumstellar
shell.  Whether the positions of the objects in the color-color
diagrams can be directly related to the temperature and core mass of
the central star needs further investigation.

\section{Conclusions}
\label{concl}

In this article we reported further investigations of the
\iras\ selected sample of PN candidates that was
presented in Van de Steene \& Pottasch (\cite{VdSteene93}). About 20~\%
of the candidates in that sample have been detected in the radio
and/or H$\alpha$ and were later confirmed as PNe. 
Here we investigated the nature of the non-radio-detected sources.

\begin{itemize}

\item Of sixteen positively identified objects, seven show \brg\ in
absorption.  The absorption lines are very narrow in six objects,
indicating a low surface gravity. This is a strong indication for the
post-AGB nature of these objects. Another six objects show \brg\
in emission. Two of these also show photospheric absorption lines.
All emission line sources have a strong underlying continuum, unlike
normal PNe.  In another three objects, no clear \brg\ absorption
or emission was visible.

\item The objects showing \brg\ in emission were re-observed in the
radio continuum with the Australia Telescope Compact Array.  None of
them were detected above a detection limit of 0.55~mJy/beam at 6~cm
and 0.7~mJy/beam at 3~cm, while they should have been easily seen if
the radio emission was optically thin and Case~B recombination was
applicable. It is argued that the \brg\ emission may be due to
ionization in the post-AGB wind, present before the star is hot enough
to ionize the AGB shell.

\item The fact that our objects were mostly selected from the region in the
\iras\ color-color diagram where typically PNe are found, may
explain our higher detection rate of emission line objects compared to
previous studies, which selected their candidates from a region
between AGB and PNe. These post-AGB stars also cover a larger range in
color and are generally much redder than the ones known so far.

\item  
In the near-infrared color-color diagrams our objects cover a very
large range of extinction.  Near-infrared versus far-infrared
color-color diagrams show trends which reflect the expected evolution
of the expanding circumstellar shell.  No distinction can be made
between the objects showing \brg\ in emission, absorption, or a flat
spectrum in the near- and far-infrared color-color diagrams.  Whether
the positions of the objects in the color-color diagrams can be
directly related to the temperature and core mass of the central star needs further
investigation.

\item We identified the \jk$-$\jl\ versus \jl$-$[25] 
diagram as a potentially useful tool to distinguish: 1) extended from
unresolved post-AGB stars, and 2) obscured objects of \class~II having
thick circumstellar shells from the brighter \class~IV objects which
show a stellar signature in their near-infrared SEDs.  However, this
result should be confirmed with a larger sample.

\end{itemize}

\begin{figure}
\caption{Countour plots of resolved objects in the \jn-band.
The positions not corrected for the pointing errors of the \eso\ 3.6-m telescope.}
\center{\epsfxsize=9cm \epsfbox{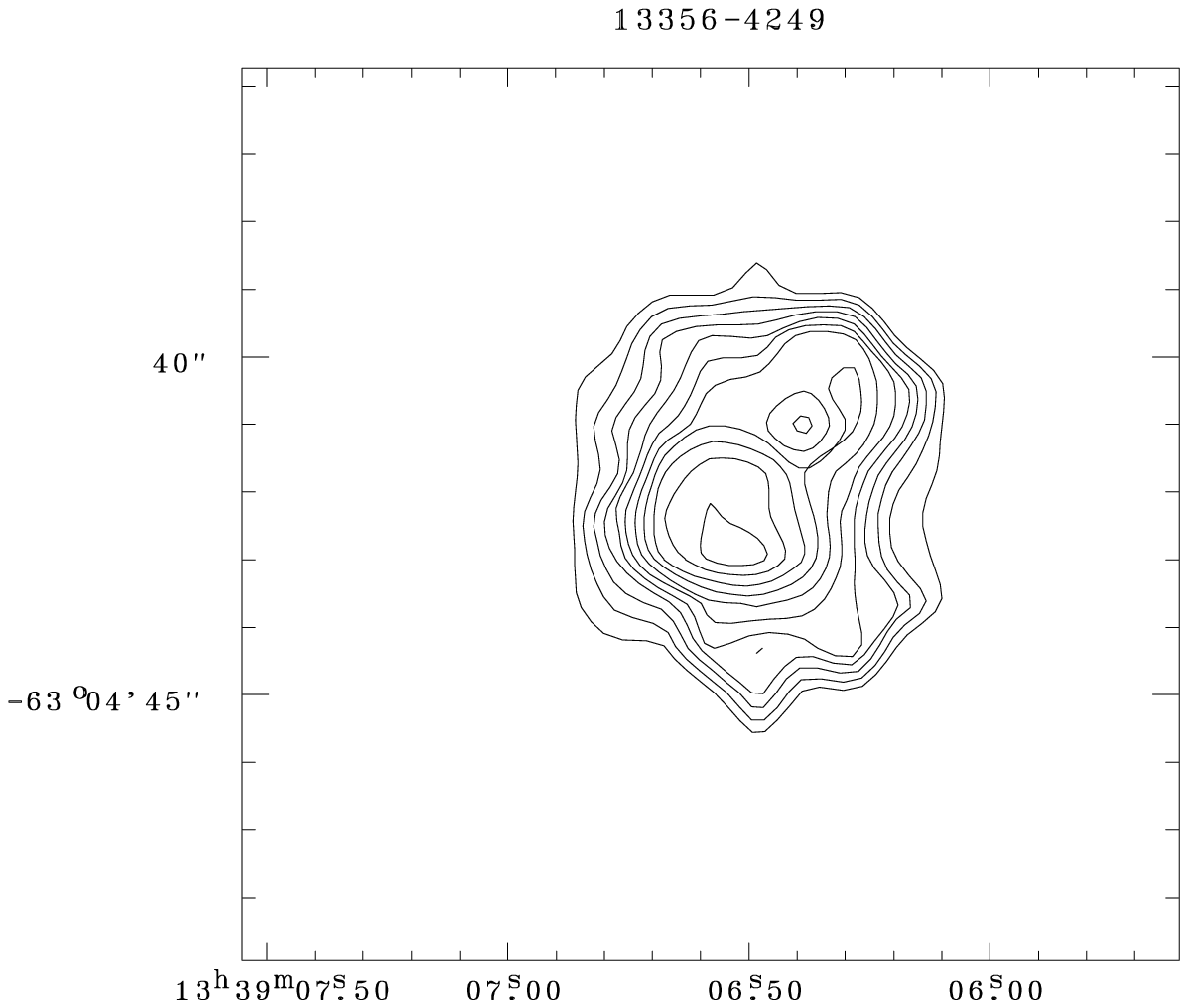}}
\center{\epsfxsize=9cm \epsfbox{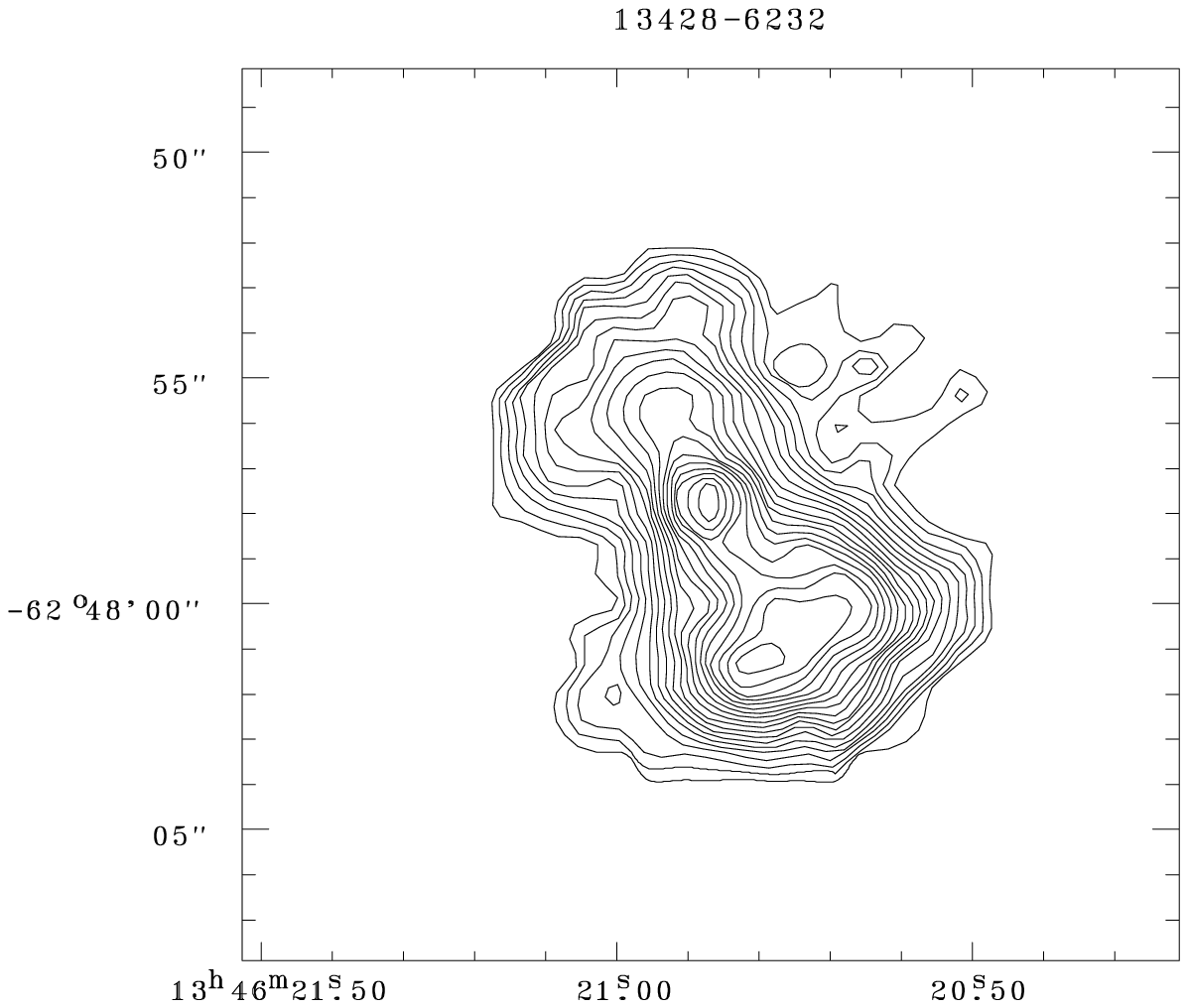}}
\center{\epsfxsize=9cm \epsfbox{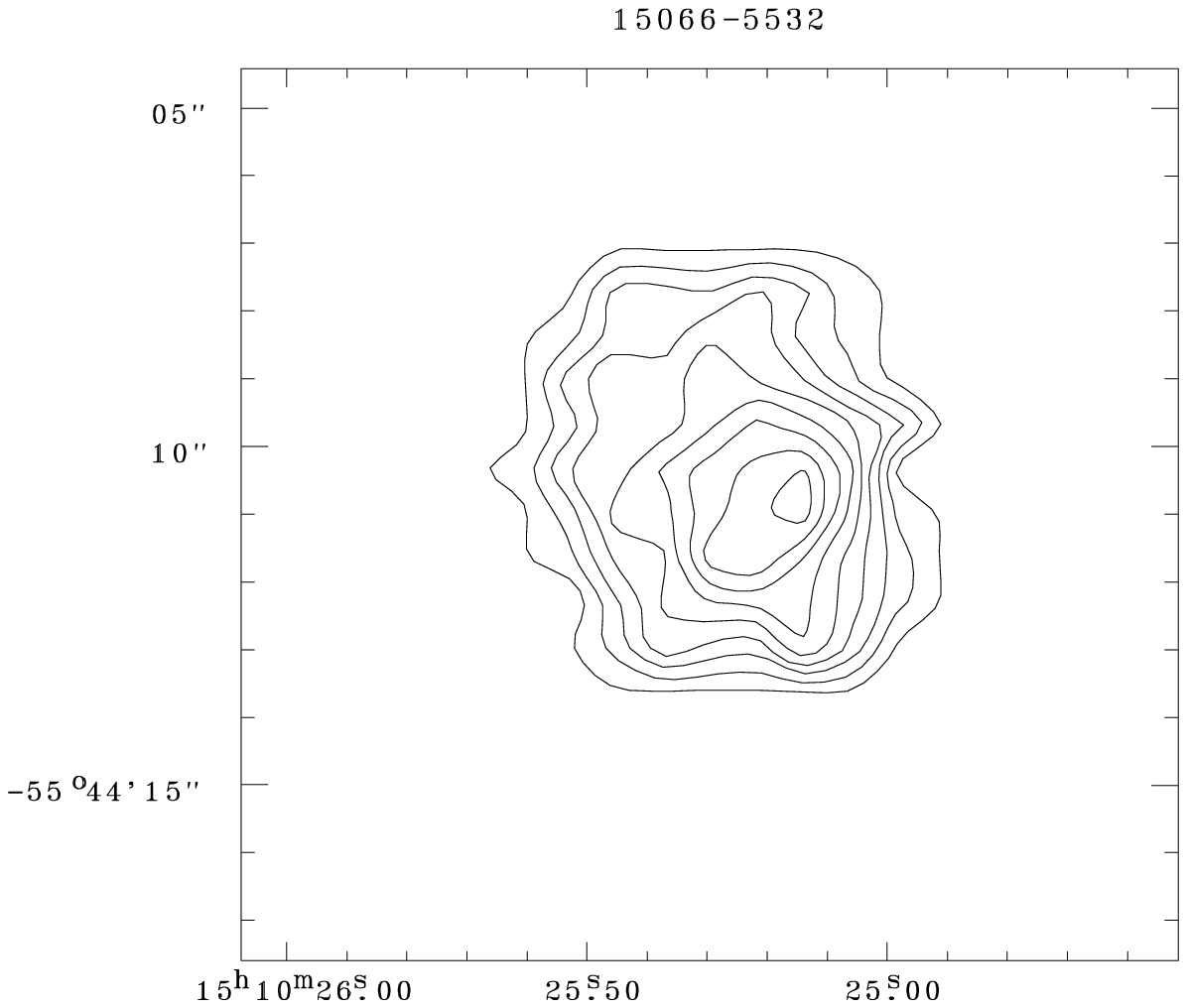}}
\label{plotsN}
\end{figure}
\begin{figure}
\center{\epsfxsize=9cm \epsfbox{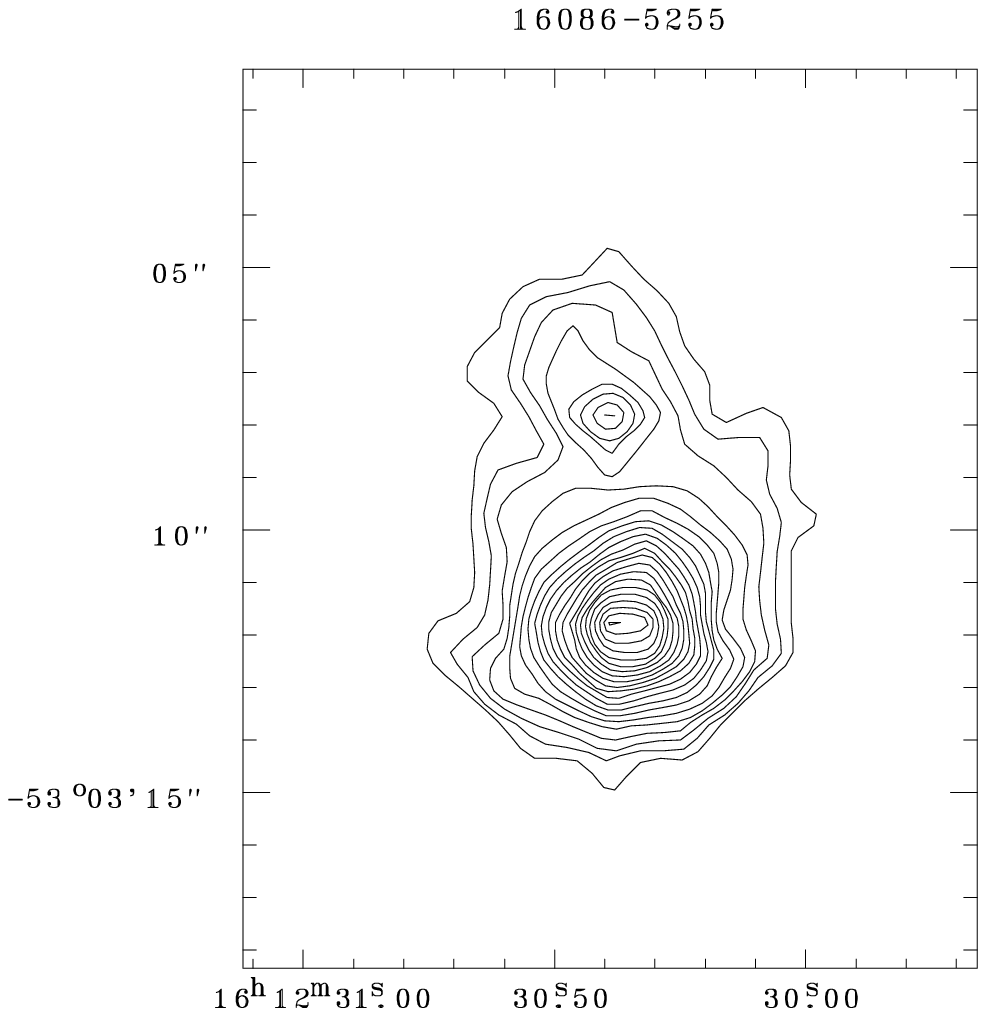}}
\center{\epsfxsize=9cm \epsfbox{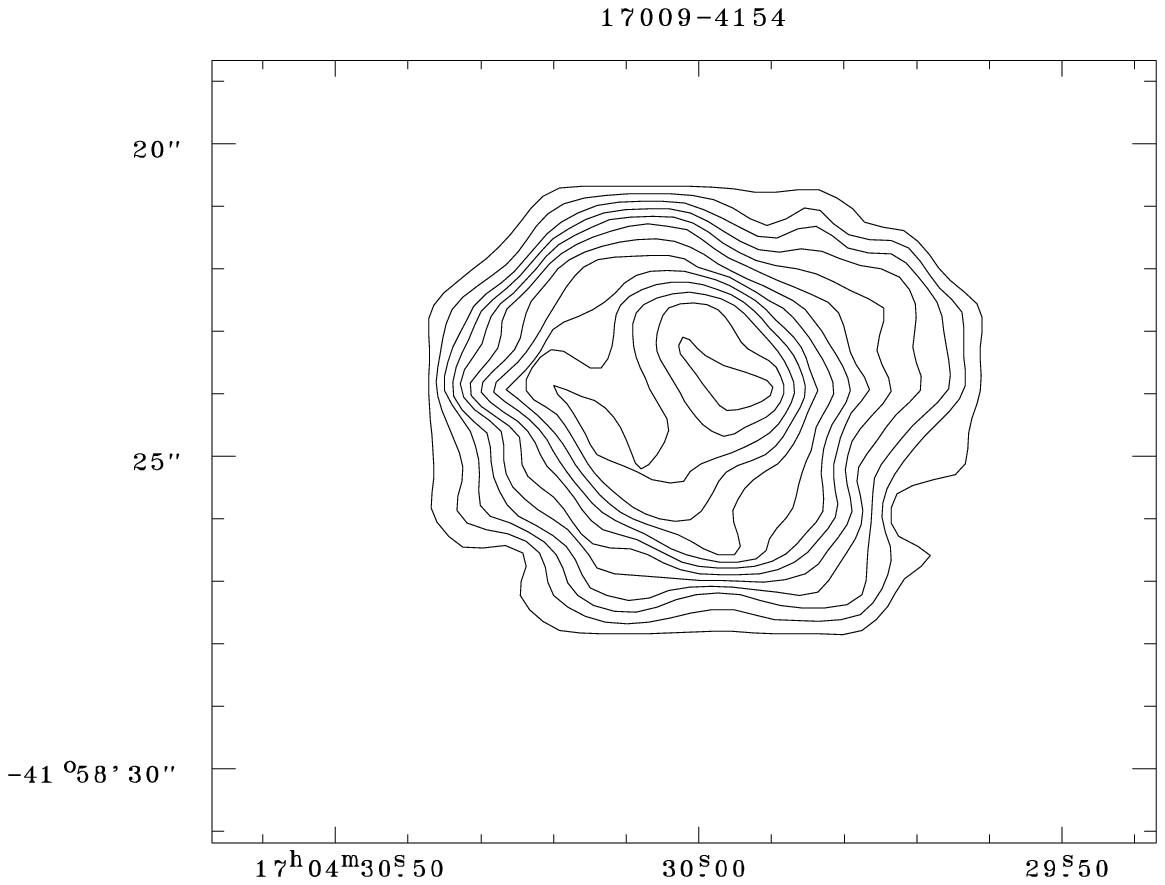}}
\end{figure}

\begin{figure}
\caption{The \caspir\ \jk-band images of the \iras\ sources.
The \iras\ position is indicated with a cross and the object observed with 
\irspec\ is in the box. The encircled object might be a more likely 
\iras\ counterpart. The pixelsize is 0\farcs25 and the 
FOV 1\arcmin $\times$ 1\arcmin. North is up and East to the left.}
\vspace*{5mm}\center{\large\bf These images are supplied as 20 separate jpeg images
(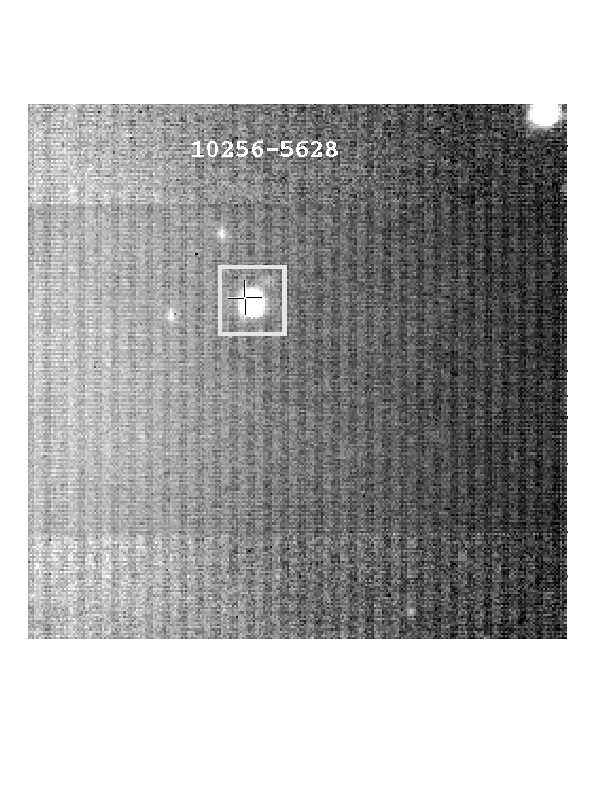 through 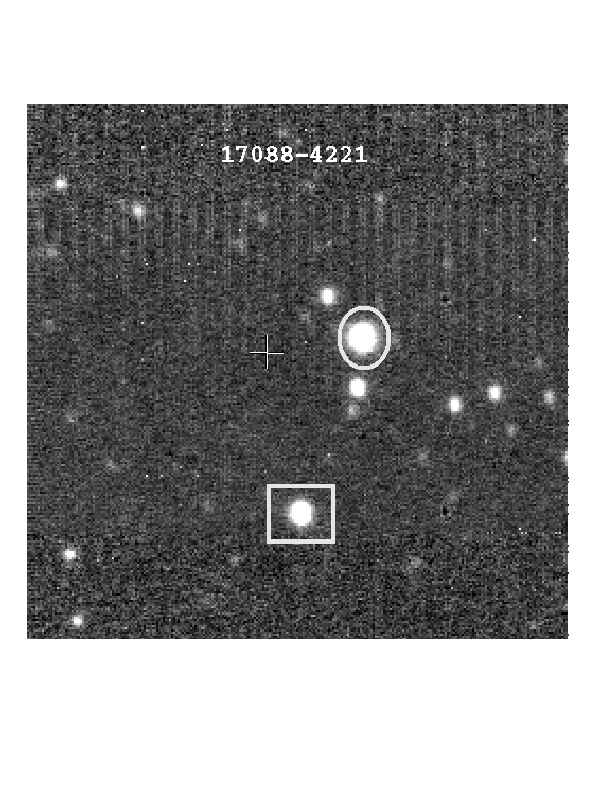).}
\label{Kimages}
\end{figure}

\begin{figure*}
\label{sedsfig}
\caption{
We present the spectral energy distribution of the sources: plotted
are the near-infrared \jhkl\ flux values at 1.29~$\mu$m, 1.65~$\mu$m,
2.20~$\mu$m, and 3.85~$\mu$m respectively, and the far-infrared \iras\
12-$\mu$m, 25-$\mu$m, and 60-$\mu$m flux values.}
\center{\epsfxsize=18cm \epsfbox{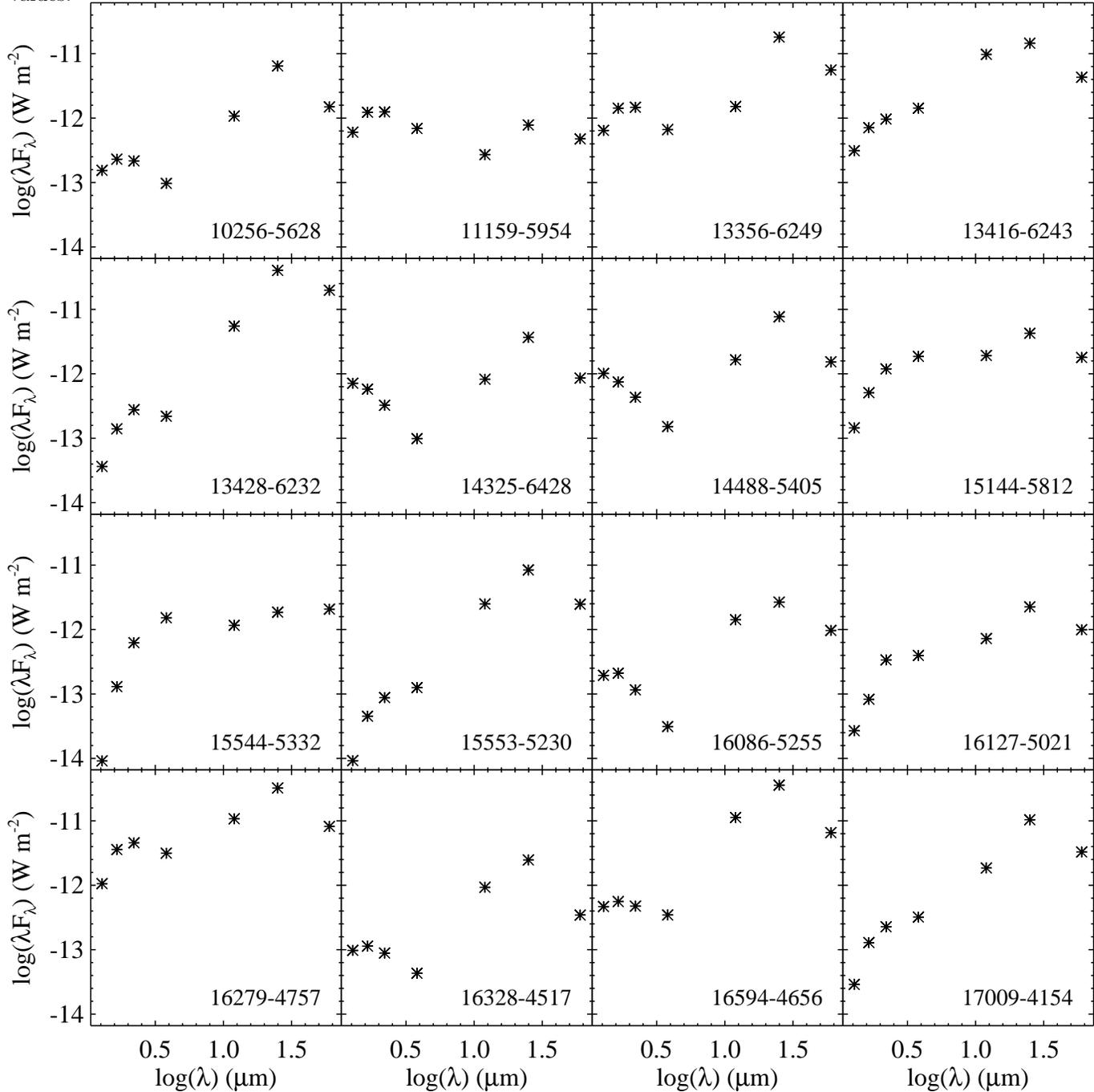}}
\end{figure*}

\begin{figure*}
\caption{The normalized spectra
of the objects with \brg\ in emission, absorption, and objects
for which no \brg\ was detected in absorption or emission.}
\center{\epsfxsize=18cm \epsfbox{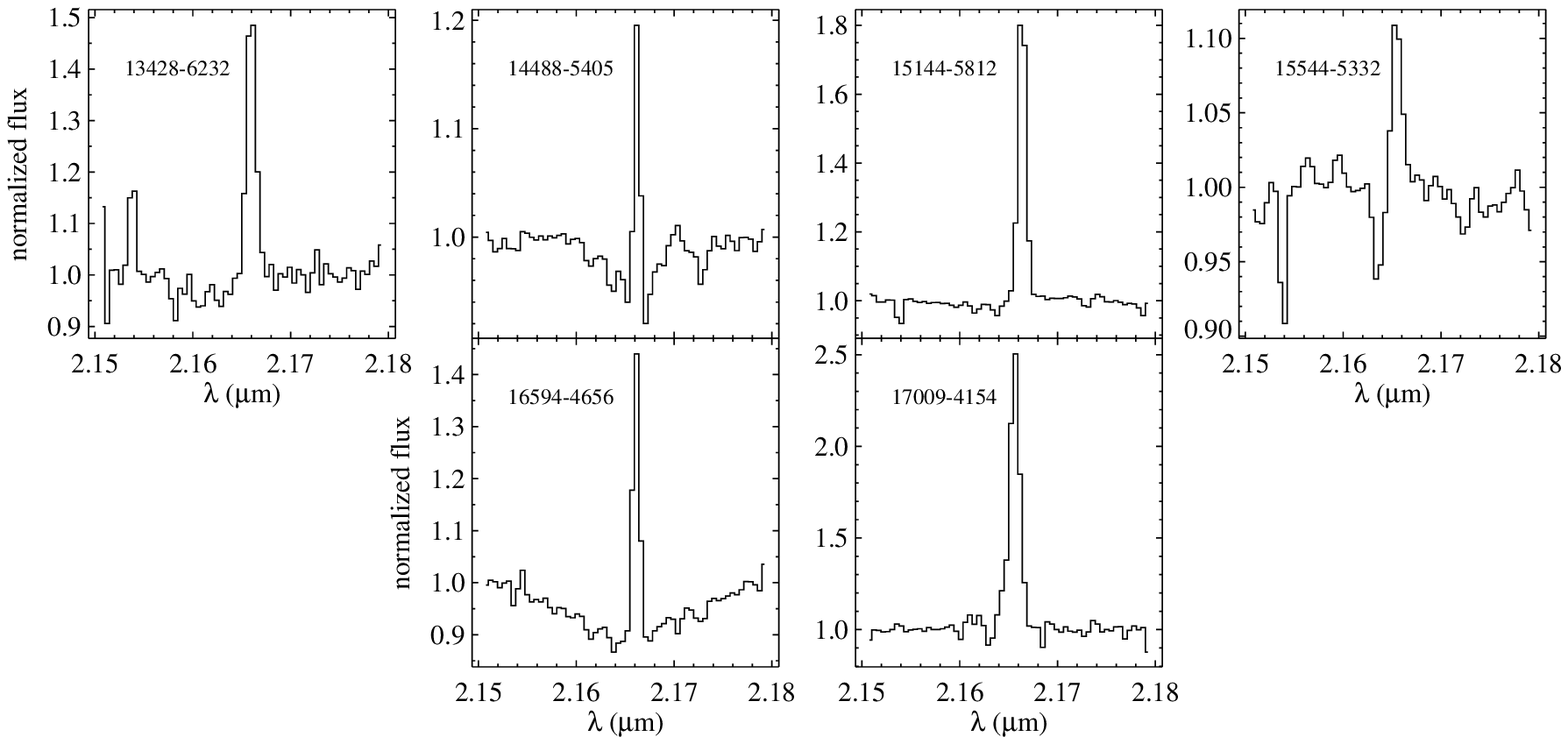}}
\vspace*{\dblfloatsep}
\center{\epsfxsize=18cm \epsfbox{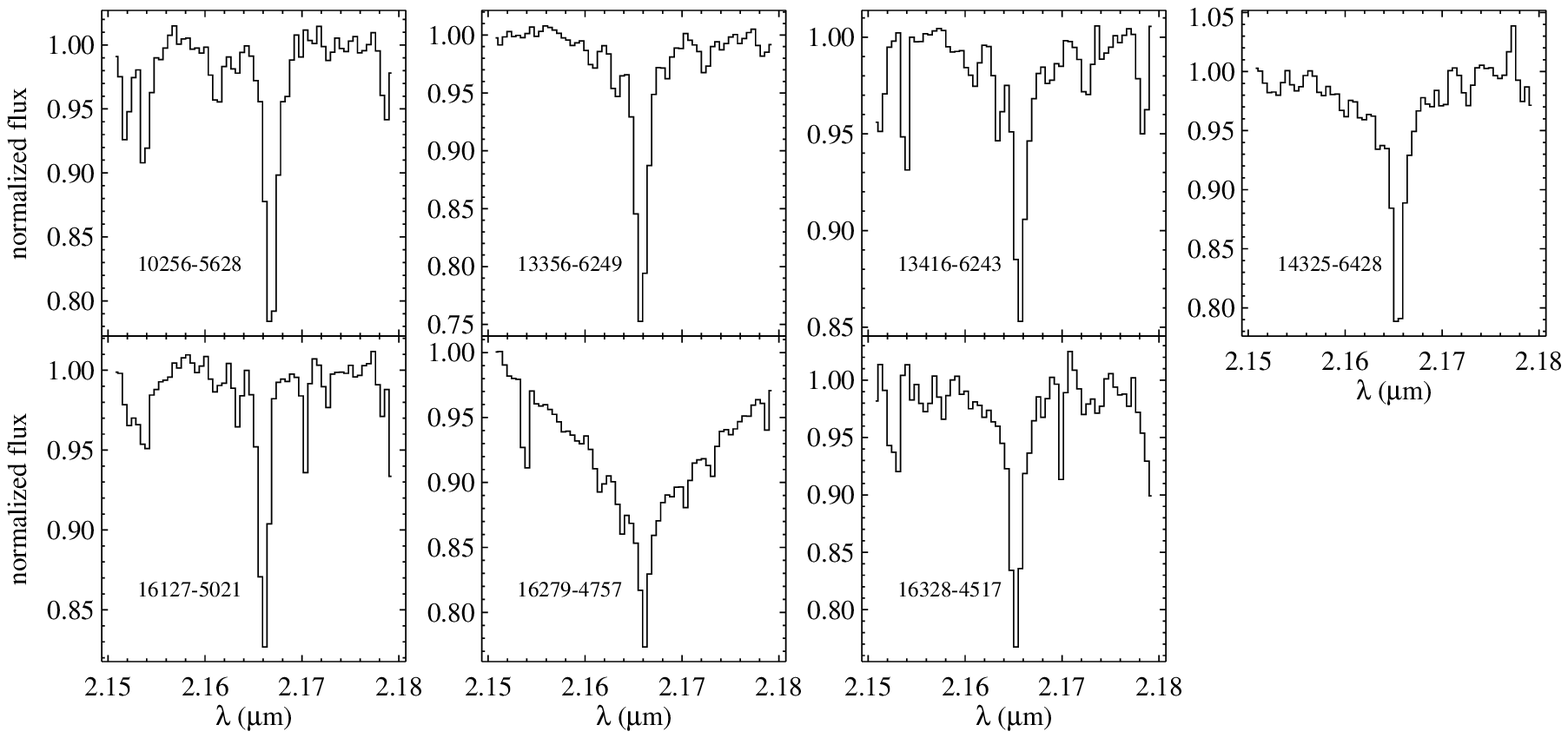}}
\vspace*{\dblfloatsep}
\center{\epsfxsize=18cm \epsfbox{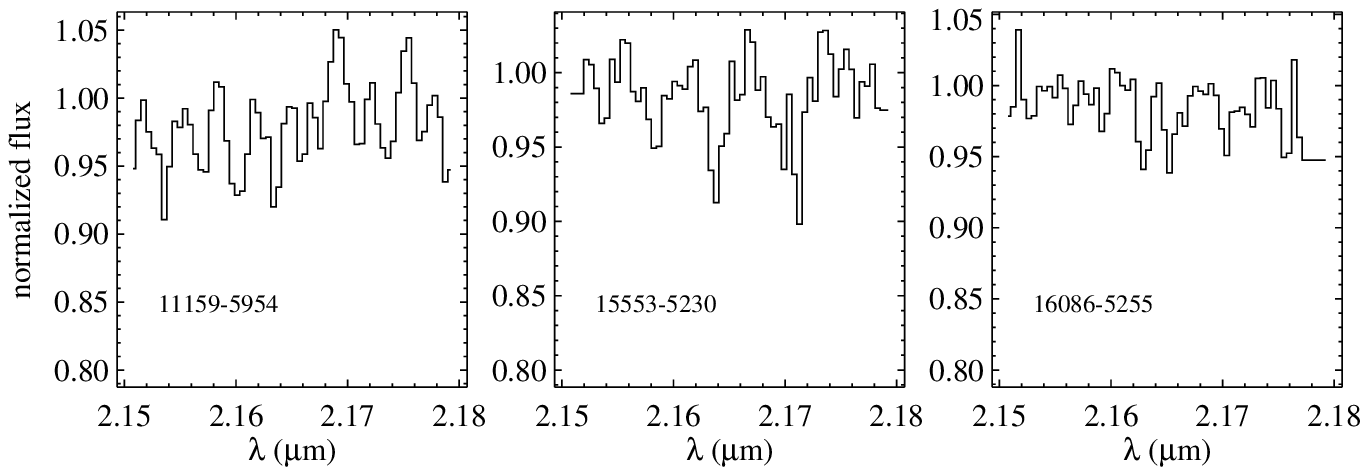}}
\label{emspec}
\end{figure*}




\begin{acknowledgements}

PvH wishes to thank ESO for their hospitality and financial support
during his stay in Santiago where part of this paper was written. PvH
acknowledges support by the Netherlands Foundation for Research in
Astronomy (ASTRON) through grant no.\ 782--372--033 during his stay in
Groningen and the NSF through grant no.\ AST 96--17083 during his stay
in Lexington. 

SkyCat was developed by ESO's Data Management and Very Large Telescope
(VLT) Project divisions with contributions from the Canadian
Astronomical Data Center (CADC).


\end{acknowledgements}


\begin{thebibliography}{}
%

\bibitem[1984]{Beichman84}
Beichman C.A., Neugebauer G., Habing H.J., Clegg P.E., Chester T.J., 1984,
IRAS Catalogs and Atlases Explanatory Supplement

\bibitem[1995]{Bloecker95}
Bl\"ocker T., 1995, A\&A 299, 755


\bibitem[1994]{Epch94}
Epchtein N., de Batz B., Copet E., et al., 1994, Ap\&SS 217, 3

\bibitem[1993]{Frank93}
Frank A., Balick B., Icke V., Mellema G., 1993, ApJ 404, L25

\bibitem[1997]{Garcia97}
Garc\'{\i}a-Lario P., Manchado A., Pych W., Pottasch S.R., 1997, A\&A 126, 479 




\bibitem[1997]{Hakkila97}
Hakkila J., Myers J.M., Stidham B.J., Hartmann D.H., 1997,
AJ 114, 2043

\bibitem[1991]{Hirsh91}
Hirshfeld A., Sinnott R.W., Ochsenbein F., 1991, Sky Catalogue 2000.0 2nd
ed., Sky Publishing Corporation, Cambridge Massachusetts

\bibitem[1999]{Hrivnak99}
Hrivnak B.J., Kwok S., Su K.Y.L., 1999, ApJ 524, 849

\bibitem[1993]{Hu93}
Hu J.Y., Slijkhuis S., de Jong T., Jiang B.W., 1993, A\&AS 100, 413

\bibitem[1993]{Kaufl93}
K\"aufl H.U., Renzini A., Stanghellini L., 1993, ApJ 410, 251

\bibitem[1995]{Knapp95}
Knapp G.R., Bowers P.F., Young K., Philips T.G., 1995, ApJ 455, 293

\bibitem[1982]{Kwok82}
Kwok S., 1982, ApJ 258, 280




\bibitem[1994]{McGregor94}
McGregor P., 1994, \caspir\ manual, http:\-//www.\-mso.\-anu.\-edu.\-au/\-ob\-ser\-ving/\-tel\-docs/\-2.3m/\-CASPIR


\bibitem[1992]{Oliva92}
Oliva E., Origlia L., 1992, A\&A 254, 466


\bibitem[1998]{Sahai98}
Sahai R., Trauger J.T., 1998, AJ 113, 1357

\bibitem[1999]{Soker99}
Soker N., 1999, MNRAS, submitted (astro-ph/9912015)

\bibitem[1994]{Solf94}
Solf J., 1994, A\&A 282, 567 

\bibitem[2000]{Ueta00}
Ueta T., Meixner M., Bobrowsky M., 2000, ApJ 528, 861

\bibitem[1988]{vdVeen88}
van der Veen W.E.J.C., Habing H.J., 1988, A\&A 194, 125

\bibitem[1989]{vdVeen89}
van der Veen W.E.J.C., Habing H.J., Geballe T.R., 1989, A\&A 226, 108

\bibitem[1993]{VdSteene93}
Van de Steene G.C., Pottasch S.R., 1993, A\&A 274, 895 

\bibitem[1995]{VdSteene95}
Van de Steene G.C., Pottasch S.R., 1995, A\&A 299, 238

\bibitem[1996b]{VdSteene96b}
Van de Steene G.C., Jacoby G.H., Pottasch S.R., 1996, A\&AS 118, 243 

\bibitem[1996a]{VdSteene96a}
Van de Steene G.C., Sahu K.S., Pottasch S.R., 1996, A\&AS 120, 111 

\bibitem[2000]{VdSteene00}
Van de Steene G.C., Wood P.R., van Hoof P.A.M., 2000,
in Kastner J.H., Soker N., Rappaport S.A. (eds.)
Asymmetrical Planetary Nebulae II: from Origins to Microstructures.
ASP Conference Series, Vol.~199, p.~191

\bibitem[1997]{vHoof97}
van Hoof P.A.M., Oudmaijer R.D., Waters L.B.F.M., 1997, MNRAS 289, 371

\bibitem[1989]{Volk89}
Volk K.M.,  Kwok S.,  1989, ApJ 342, 345  


\end{thebibliography}
\end{document}